\documentclass[aps,11pt,prx,longbibliography]{revtex4-1}
\usepackage[latin1]{inputenc} 
\usepackage[T1]{fontenc}
\usepackage{lmodern}
\usepackage{graphicx}
\usepackage{amsmath,amssymb,color,xcolor,bm}
\renewcommand{\d}{\mathrm{d}}
\newcommand{\rr}{\mathbf{r}}
\renewcommand{\v}{\mathbf{v}}
\newcommand{\q}{\mathbf{q}}
\newcommand{\rev}[1]{{\color{black} {#1}}}

\begin{document}
\title{Fluctuation-induced quantum friction in nanoscale water flows}

\author{Nikita Kavokine$^1$, Marie-Laure Bocquet$^2$ and Lyd\'eric Bocquet$^1$}
\email{e-mail: nikita.kavokine@ens.fr; lyderic.bocquet@ens.fr}
\affiliation{$^1$Laboratoire de Physique de l'\'Ecole Normale Sup\'erieure, ENS, Universit\'e PSL, CNRS, Sorbonne Universit\'e, Universit\'e Paris-Diderot, Sorbonne Paris Cit\'e, Paris, France}
\affiliation{$^2$PASTEUR, D\'epartement de Chimie, \'Ecole Normale Sup\'erieure, PSL University, Sorbonne Universit\'es, CNRS, 24 Rue Lhomond, 75005, Paris, France}

\begin{abstract}
\textbf{
The flow of water in carbon nanochannels has defied understanding thus far \cite{Bocquet2020}, with accumulating experimental evidence for ultra-low friction, exceptionally high water flow rates, and curvature-dependent hydrodynamic slippage~\cite{Holt2006,Whitby2008,Secchi2016,Duan2018}.
These unique properties have raised considerable interest in carbon-based membranes for desalination, molecular sieving and energy harvesting \cite{Park2014a,Joshi2014}. However, the mechanism of water-carbon friction remains unknown \cite{Faucher2019}, with neither current theories~\cite{Bocquet2007}, nor
classical \cite{Thomas2008,Falk2010} or \emph{ab initio} molecular dynamics simulations \cite{Tocci2014} providing satisfactory rationalisation for its singular behaviour.
Here, we develop a quantum theory of the solid-liquid interface, which reveals a new contribution to friction, due to the coupling of charge fluctuations in the liquid to electronic excitations in the solid. We expect that this quantum friction, which is absent in {Born-Oppenheimer molecular dynamics}, is the dominant friction mechanism for water on carbon-based materials. {As a key result}, we demonstrate a {dramatic} difference in quantum friction between the water-graphene and water-graphite interface, due to the coupling of water Debye collective modes with a thermally excited plasmon specific to graphite. 
This suggests an explanation for the radius-dependent slippage of water in carbon nanotubes~\cite{Secchi2016}, in terms of the nanotubes' electronic excitations. 
Our findings open the way to quantum engineering of hydrodynamic flows through the confining wall electronic properties.
}
\end{abstract}


\maketitle

\rev{While the liquid flow rate through a macroscopic channel is determined solely by its geometry, the permeability of nanoscale channels depends strongly on the amount of liquid friction at the channel walls~\cite{Kavokine2021}}. Yet, understanding liquid friction on solid surfaces remains a key fundamental challenge in fluid dynamics~\cite{Faucher2019}, with the water-carbon interface presenting a particularly puzzling picture~\cite{Bocquet2020}. Water was indeed found to exhibit very strong slippage -- that is, low friction -- in carbon nanotubes (CNTs)~\cite{Holt2006}, with the slippage increasing with decreasing tube radius~\cite{Secchi2016}. Yet, almost no slippage was detected inside structurally similar, but electronically different, boron nitride nanotubes (BNNTs), and only moderate water slippage has been reported on flat graphite surfaces~\cite{Maali2008}.
These observations challenge the current theories of the solid-liquid interface, which are based on picturing the solid as a static external potential that acts on the fluid molecules~\cite{Barrat1999}. In this picture, friction results merely from surface roughness, as flows induced on the roughness scale dissipate mechanical energy. 
However, this classical approach involves a high degree of arbitrariness when applied to atomically smooth surfaces, through the choice of the molecular force fields. It leads, for instance, to a three order of magnitude spread in the reported molecular dynamics (MD) simulation results for the water friction coefficient inside sub-10 nm CNTs~\cite{Sam2020}. Moreover, it fails to account for certain experimental observations : in particular, the radius-dependence of friction observed for relatively large {(over $60~\rm nm$ in diameter)} \rev{multiwall} CNTs~\cite{Secchi2016}, or the {dramatic} difference with BNNTs.
All this suggests that some key ingredients are missing from the current understanding of the solid-liquid interface, which might include quantum effects in nanoscale fluid dynamics. 
A few simulation studies have made pioneering attempts at taking into account electronic degrees of freedom at the solid-water interface, through polarisable force fields~\cite{Misra2017} or \emph{ab initio} molecular dynamics (AIMD)~\cite{Tocci2014} in the Born-Oppenheimer approximation. While the latter point to some difference in water friction between graphene and boron nitride surfaces, such simulations are still insufficient to account for the whole experimental picture. 

In this Article, we introduce a theoretical description of the solid-liquid interface that fully accounts for the quantum dynamics of the {electrons in the solid}. We predict that quantum effects do contribute to the solid-liquid friction, with the water-carbon interface being unique in many respects.

\vskip0.5cm
\noindent{\bf \large Single particle friction} \\
In order to introduce the theoretical framework, we first consider a single point charge $e$ moving at a height $h$ parallel to a solid surface lying in the $(x,y)$ plane. 
As electronic degrees of freedom are taken into account, a polarisation charge arises within the solid, and dissipation then occurs in the dynamics of this polarisation charge.  
This mechanism is best formalised by introducing the solid surface response function $g_{\rm e} (\q,\omega)$, which 
is defined in terms of the density-density response function $\chi_{\rm e}$: 
\begin{equation}
g_{\rm e}(\q,\omega) = \frac{-e^2}{4 \pi \epsilon_0 }\frac{2\pi}{q} \int_{-\infty}^0 \d z \d z' e^{q(z+z')} \chi_{\rm e}(\q,z,z',\omega),
\label{defg}
\end{equation}
{with the wavevector $\q$ lying in the plane of the interface}. Physically, $g_{\rm e} (\q,\omega)$ relates the external potential applied to the solid in the half-space $z<0$ to the induced potential in the half-space $z>0$. If the external potential is an evanescent plane wave of the form $V_{\rm ext} = V_0 e^{i(\q \rr - \omega t)}e^{qz}$, the induced potential is $V_{\rm ind} = V_0 g_{\rm e}(\q,\omega) e^{i(\q \rr - \omega t)}e^{-qz}$ (see discussion in the Supplementary Information (SI) section 3.1). 
The friction force on the charge moving at velocity $\v$ then writes (SI section 1) 
\begin{equation}
\mathbf{f} = \frac{-e^2}{8 \pi^2 \epsilon_0} \int \d   \q  \, \frac{\q}{q} e^{-2qh} \,  \mathrm{Im}\,  g_{\rm e}(\q, \q \cdot \v).
\label{particle_force}
\end{equation}
Eq.~\eqref{particle_force} accounts for electronic friction, that is, friction through the generation of electronic excitations within the solid. 
This phenomenon has been invoked in various situations where classical nuclear degrees of freedom are coupled to an electron bath~\cite{Wodtke2004,Dou2018}.

The mechanism outlined here for a single charged particle should in principle apply as well to a dense polar liquid such as water. 
An electronic contribution to hydrodynamic friction, given by the sum of electronic friction forces on each water molecule, was indeed proposed in \cite{Sokoloff2018} to account for the radius-dependent slippage in carbon nanotubes, but the prediction disagreed with experiments by orders of magnitude. The pitfall in these estimates lies in the wrong treatment of the collective excitations of the fluid, which cannot be accounted for by a sum of individual contributions.

\begin{figure}
\centering
\includegraphics{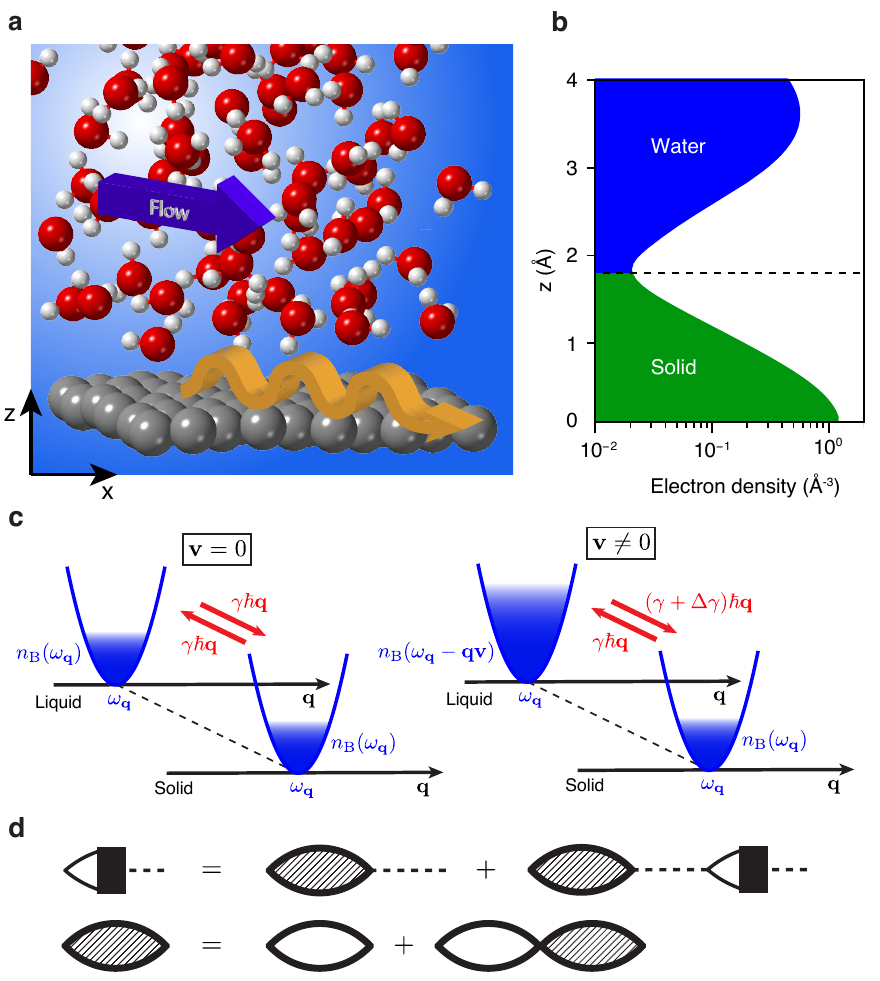}
\caption{\textbf{Quantum friction at the solid-liquid interface}. \textbf{a}. Artist's view of the quantum friction phenomenon: water charge fluctuations couple to electronic excitations within the solid surface, represented by the orange arrow.  \textbf{b}. Average electronic density, as obtained from density functional calculations (SI, section 7), at the water-graphene interface. \textbf{c}. Schematic of the quantum friction mechanism, showing quasiparticle tunnelling between two surface modes at wavevector $\q$ and frequency $\omega_{\q}$. The filling of the blue parabolas represents the occupation of each mode, according to the Bose-Einstein distribution $n_{\rm B}$. The back and forth tunnelling rates $\gamma$ are different in the presence of flow, resulting in net momentum transfer from the liquid to the solid. Further details are given in the SI, section 2.8. \textbf{d}. Feynman diagram representation of the Dyson equation for the electron-water density correlation function. Full lines are electron propagators, and dashed lines are water propagators. {The equation expresses that electron-water correlations are mediated by all possible coupled fluctuations of the water and electron densities.}}
\end{figure}

\newpage
\vskip0.5cm
\noindent{\bf \large Many-body interfacial dissipation}\\
We are hence required to go beyond the single-particle picture. As a first step, we formally express the friction force acting on a liquid flowing on a solid surface in terms of the system's many-body dynamics. 
To do so, we build on the general framework of fluctuation-induced forces~\cite{Volokitin2007}, which is specifically extended to account for a solid-liquid instead of a solid-solid interface. 

Let us assume for concreteness that the liquid is water. 
We focus on the force due to the long range Coulomb interaction between the water molecules and the solid electrons, noting that we could consider the Coulomb interactions between water and the crystal lattice in a similar way. Since the dynamics of the electrons are quantum, the friction force is represented by an operator $ \mathbf{\hat F}$, whose average value at time $t$ is 
\begin{equation}
\langle \mathbf{ \hat  F}(t) \rangle = -\int \d \rr\, \d \rr' \, \nabla_{\parallel}V (\rr - \rr') \langle n_{\rm w}(\rr',t) \hat n_{\rm e} (\rr,t)\rangle.
\label{force_micro}
\end{equation}
Here $V$ is the Coulomb potential, $n_{\rm w}$ is the instantaneous charge density of water, $\hat n_{\rm e}$ is the electron density operator, and $\nabla_{\parallel}$ represents the gradient parallel to the interface; the average is over all thermal and quantum fluctuations of the system. 
While the dynamics of water are well described within classical mechanics even at the molecular scale, the electron dynamics are intrinsically quantum. Hence, to ensure a consistent formalism, we represent the water charge density by a gaussian quantum field $\hat n_{\rm w}$ with prescribed correlation functions, as is done, for example, in the theory of solvation~\cite{Song1996}. The average in Eq.~\eqref{force_micro} is then computed in the framework of many-body perturbation theory, with respect to the interaction Hamiltonian that comprises both the electron-water and electron-electron Coulomb interactions: 
\begin{equation}
\hat H_{\rm int} = \int \d \rr \d \rr' \hat n_{\rm w}(\rr',t) V(\rr- \rr') \hat n_{\rm e} (\rr,t) + \frac{1}{2} \int \d \rr \d \rr' \hat n_{\rm e}(\rr',t) V(\rr-\rr') \hat n_{\rm e}(\rr,t). 
\label{hint}
\end{equation}

The main difficulty of the computation is that it deals with a non-equilibrium steady state: for the friction force to be non-zero, we need to impose a net flow of water parallel to the interface. 
Accordingly, we treat the perturbative expansion in the non-equilibrium Schwinger-Keldysh framework~\cite{Rammer1986}. 
After an exhaustive computation, reported in SI section 2, our most general result is a Dyson equation relating the water-electron, water-water and electron-electron density correlation functions, whose Feynman diagram representation is given in Fig.~1d. The friction force may then be expressed in terms of the Keldysh component of the water-electron correlation function. 
We stress here the generality of the result, as it applies to a fully out of equilibrium situation, and allows for overlap between the water and electron densities. Moreover, we treat the water-electron interaction beyond the mean-field approximation, as our computation allows for a self-energy correction due to the presence of water in the electron Green's functions. 

The fully general but cumbersome Dyson equation can be simplified if the water and electron densities are no longer allowed to interpenetrate each other. This is a reasonable approximation as long as there is no chemisorption of water on the solid surface: then, the short-range Pauli repulsion effectively acts as an infinite potential barrier between the water and the electrons. Density functional calculations show that this is the case, for instance, at the water-graphene interface (Fig.~1b).  
The hydrodynamic flow profile above the solid surface is assumed to be uniform and equal to the interfacial velocity $\v$, which is justified
as long as the typical range of the solid-liquid interactions is smaller than the slip length.
Then, expanding the Dyson equation to linear order in $\v$, we obtain a closed expression of the form $\langle \mathbf{ \hat  F} \rangle /\mathcal{A}= -\lambda \mathbf{v}$, where $\mathcal{A}$ is the surface area, and $\lambda$ is the solid-liquid friction coefficient. It separates into two terms: $\lambda = \lambda_{\rm Cl} + \lambda_{\rm Q}$, with
\begin{equation}
\lambda_{\rm Cl} = \frac{1}{4 \pi^2 k_{\rm B} T} \int \d \q \, \frac{(\q \cdot \v)^2}{v^2} |V_{\rm e}(\q)|^2 \int_{0}^{+\infty} \d t \, S_{\rm w} (\q,t)
\label{Clfriction}
\end{equation}
and
\begin{equation}
v \cdot \lambda_{\rm Q} = \frac{1}{8 \pi^2} \int_0^{+\infty}q \mathrm{d} q \, (\hbar q)\int_0^{+\infty} \frac{\mathrm{d}( \hbar \omega)}{k_B T} \frac{ q\cdot v }{ \mathrm{sinh}^2(\frac{\hbar \omega}{2k_{\rm B} T})}\frac{ \mathrm{Im}[g_{\rm e}(q,\omega)] \, \mathrm{Im}[g_{\rm w}(q,\omega)] }{|1-g_{\rm e}(q,\omega)\,g_{\rm w}(q,\omega)|^2}.
\label{qfriction}
\end{equation}
We refer to the first contribution $\lambda_{\rm Cl}$ in Eq.~\eqref{Clfriction} as the classical term, as it exactly reproduces the result obtained in the conventional surface roughness picture~\cite{Barrat1999,Falk2010}.  
Here, $V_{\rm e}(\q)$ is the average Coulomb potential acting on the interfacial water layer, and $S_{\rm w}(\q,t)$ is the water charge structure factor (precise definitions of these quantities are given in the SI, section 2.5). 

The second contribution $\lambda_{\rm Q}$ in Eq.~\eqref{qfriction} is absent in the roughness-based picture, and originates from the coupled water and electron dynamics, as it involves the surface response functions of both the solid and the liquid, $g_{\rm e}$ and $g_{\rm w}$, respectively. 
We call it the quantum friction term. 
A similar term is known to arise in the non-contact friction between two dielectric media separated by a vacuum gap~\cite{Volokitin2007}: it is then interpreted as a dynamic analogue of the van der Waals force. However, the expression of this van der Waals friction was rigorously established only in the case of two media with local dielectric response~\cite{Volokitin2006}, hence it may not apply directly to the solid-liquid interface under scrutiny. 

\rev{Since it is accessible through MD simulations, the classical contribution has been amply studied for numerous solid-liquid systems~\cite{Huang2008}. We will therefore discuss in detail only the quantum contribution in Eq.~\eqref{qfriction}, which has not been considered in the context of the solid-liquid interface. The structure of Eq.~\eqref{qfriction} can be understood in terms {of} quasiparticle tunnelling between the surface fluctuation modes of the two media, as detailed in the SI, section 2.8. Briefly, the quantum friction force $v \cdot\lambda_{\rm Q}$ decomposes into a (continuous) sum over the wavevectors $q$ of surface modes: $v \cdot \lambda_{\rm Q} = (1/4\pi)\int \d \q \, f_q$. The elementary friction force $f_q$ is given by the elementary momentum $\hbar q$, multiplied by an integral over the frequencies $\omega$, which plays the role of a tunnelling rate between modes at wavevector $q$. This tunnelling rate is non-zero only in the presence of a water flow, which effectively increases the occupation of the liquid modes with respect to the solid modes, as depicted schematically in Fig. 1c. We note that the contribution in Eq.~\eqref{qfriction} requires thermal fluctuations, as it vanishes at 0 temperature. A purely quantum contribution to non-contact friction, that survives at 0~K, may be derived~\cite{Pendry1997}, but it scales as $v^3$ and is hence negligible for our purposes. }

In order to proceed with quantitative evaluation of water-solid quantum friction coefficients, we need to determine the surface excitation spectra $\mathrm{Im} \, g_{\rm w,e} (q,\omega)$ of water, and of the electronic system under consideration. \rev{The hyperbolic sine in the denominator of the expression for $\lambda_{\rm Q}$ (Eq.~\eqref{qfriction}) strongly attenuates frequencies above $\omega \sim 2k_BT/\hbar$, so that mainly low energy modes below $50~\rm meV$ or 10~THz contribute to quantum friction.} However, both excitation spectra need to be evaluated up to the highest momenta, since the contribution of modes at wavevector $q$ to the quantum friction coefficient scales as $q^3$. 

\vskip0.5cm
\noindent{\bf \large Water fluctuations}\\
Our result in Eq.~\eqref{qfriction} involves the bare surface response function of water, which depends on the water dynamics in the absence of coupling to electronic degrees of freedom. Moreover, in the frequency range under consideration, the water dynamics are completely classical, so that classical MD simulations are well-suited for the determination of water surface response functions. 
Accordingly, we have carried out such simulations for water (in the SPC/E model) in contact with graphite surfaces, for which we considered two different sets of Lennard-Jones parameters (Fig. 2a). The surface response functions were determined from the equilibrium charge correlation functions through the fluctuation-dissipation theorem, according to the definition in Eq.~\eqref{defg}. Details of the simulations and analysis are given in the SI, section 4.

\begin{figure}
\centering
\includegraphics{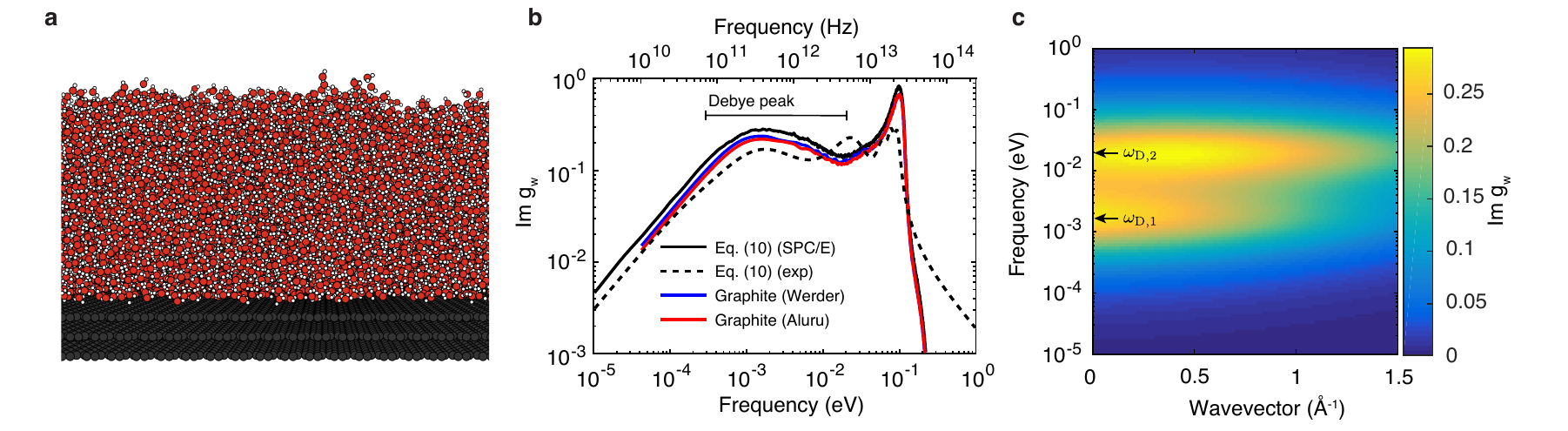}
\caption{\textbf{Surface dielectric response of water}. \textbf{a}. Snapshot of the MD simulation used for determining the water surface response function. {The graphene supercell size is $128\times 123 ~\rm \AA^2$.}  \textbf{b}. Surface response function $g_{\rm w}$ of water versus frequency, in the long wavelength limit {($q\rightarrow 0$)}. {The various curves correspond to results obtained from MD simulations of the water-graphite interface with two different sets of molecular parameters (named 'Aluru' and 'Werder', see text), and to the $q = 0$ prediction of Eq.~(\ref{gwbulk}), obtained from the experimental and simulated bulk dielectric constant.} All the determinations of the surface response function agree well in the long wavelength limit. \textbf{c}. Surface response function of water in energy-momentum space, as obtained by fitting the simulation data with two Debye peaks (Eq.~\eqref{gwfit}). }
\end{figure}

We first focus on results in the long wavelength limit ($q \to 0$), displayed in Fig. 2b. The surface response function should then converge to a value determined only by the bulk water dielectric permittivity $\epsilon_{\rm w}(\omega)$:
\begin{equation}
g_{\rm w}(0,\omega) = {\epsilon_{\rm w}(\omega)-1\over \epsilon_{\rm w}(\omega) + 1}.
\label{gwbulk}
\end{equation}
It is then interpreted as an electromagnetic reflection coefficient (see SI, section 3.1). We find that this limit is essentially reached for the lowest $q$ values accessible in our simulation ($q = 0.05~\rm \AA^{-1}$). Moreover, for frequencies below 100 meV, our results agree well with the long wavelength surface response function obtained from the experimentally determined $\epsilon_{\rm w} (\omega)$ (see SI section 4). 

The relation with the bulk dielectric function in Eq.~\eqref{gwbulk} implies that the features in $g_{\rm w} (q,\omega)$ can be analysed in terms of the well-known features of the bulk water dielectric response~\cite{Carlson2020}. 
The low-energy surface response of water is dominated by the Debye mode, which corresponds to the wide feature in the spectrum, 
spanning about three decades in frequency between 0.1~meV and 100~meV. \rev{This mode results from the collective relaxation of molecular dipoles and it is a general feature of polar liquids, so that it is also observed in polar organic solvents~\cite{Sato2004} and room-temperature ionic liquids~\cite{Koeberg2007}, albeit with lower oscillator strength than in water. Hence, our subsequent discussion, although focussed on water, applies qualitatively to quantum friction in these liquids as well. }
The sharp peak at around 100 meV corresponds to the libration mode, which involves rotation of the water molecules without displacement of their centre of mass~\cite{Carlson2020}. 

The water surface response function shows little dispersion as the momentum $q$ is increased (Figs. 2c and S1), and only small variations are found between the two models for the graphite surfaces. At large momenta ($q \geq 1~\rm \AA^{-1}$), the surface response function shows an exponential decrease, which we attribute to the depletion of water near the hydrophobic surface. We find that, for the purposes of calculation, $g_{\rm w}(q,\omega)$ is well represented by the sum of two Debye peaks at $\omega_{\rm D,1} = 1.5~\rm meV$ and $\omega_{\rm D,2} = 20 ~\rm meV$, with momentum-dependent oscillator strengths: 
\begin{equation}
g_{\rm w}(q,\omega) =  \frac{f_1(q)}{1-i \omega/\omega_{\rm D,1} } + \frac{f_2(q)}{1-i \omega/\omega_{\rm D,2} }. 
\label{gwfit}
\end{equation}
Analytical expressions for $f_1(q)$ and $f_2(q)$ are given in the SI, section 4.3; at large $q$, $f_1(q) + f_2(q) \propto e^{-2qd}$, with $d = 0.95~\rm \AA$. \rev{Further MD simulations show that the features of the water surface response relevant for quantum friction are largely unaffected by planar confinement down to 1~nm (SI section 4.4), so that the expression in Eq.~\eqref{gwfit}, determined for a semi-infinite liquid, can be safely applied to nano-confined geometries. }

\vskip0.5cm
\noindent{\bf \large Electronic excitations: the jellium model}\\
We now examine the surface excitations in the solid and first explore a generic electronic system. 
A solid has low energy electronic excitations if it {contains free charge carriers}, but the precise location of these excitations in energy-momentum space depends strongly on the exact band structure. The simplest way to account for this dependence for a wide range of parameters is in the framework of the jellium model. In the jellium model, the nuclei and core electrons are assimilated to a semi-infinite positive background, while the conduction electrons behave as free electrons (Fig. 3a), so that they occupy a parabolic band: $E(k) = \hbar^2 k^2/(2m^{*})$. The electronic structure is then completely determined by two parameters: the effective mass $m^*$ appearing in the electron dispersion, and the Fermi energy $E_{\rm F}$ up to which the band is filled. 

\begin{figure}
\centering
\includegraphics{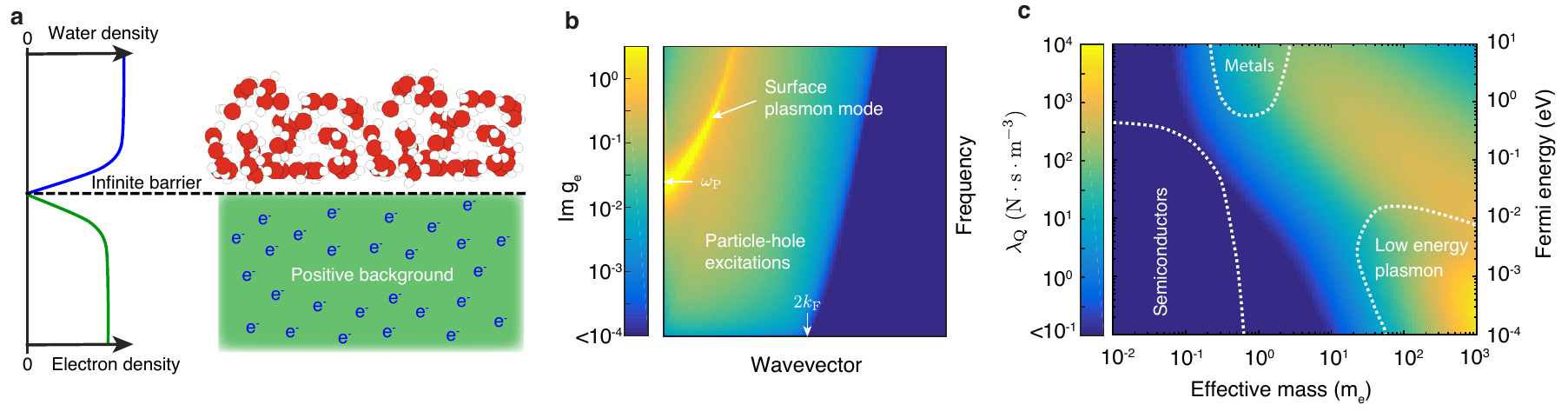}
\caption{\textbf{Quantum hydrodynamic friction on a jellium surface}. \textbf{a}. Schematic of the infinite barrier jellium model: {the water and solid electrons are separated by an infinite potential barrier}. \textbf{b}. Surface response function for a semi-infinite jellium (electron density parameter $r_{\rm s} = 5$) in energy-momentum space. \textbf{c}. Quantum friction coefficient for water on a jellium surface, as a function of the jellium Fermi energy and effective mass (in units of electron mass).}
\end{figure}

In general, in the jellium model, electrons are allowed to spill over the positive background edge~\cite{Lang1970}. However, in the presence of water, the spill-over is limited by the Pauli repulsion between the water and the surface electrons (Fig. 1b), so that the infinite barrier jellium model appears better suited to describing the electronic system under scrutiny. The infinite barrier jellium presents furthermore a technical advantage, as it may be treated within the specular reflection (SR) approximation -- where the surface response can be expressed in terms of the bulk dielectric response -- with only small quantitative differences with respect to the exact semi-infinite computation (see \cite{Penzar1984} and SI sections 3.2 and 5.1). 

A typical result for the jellium surface response function {$g_{\rm e} (q,\omega)$} in the SR approximation is shown in Fig. 3b. It presents two types of features: incoherent particle-hole excitations, and a collective surface plasmon mode. At zero energy, particle-hole excitations are present up to $q = 2k_{\rm F}$, with $k_{\rm F} = \sqrt{2 m^* E_{\rm F}/\hbar}$ the Fermi momentum. The long wavelength surface plasmon frequency $\omega_{\rm P}$ scales as $(m^*)^{1/4} E_{\rm F}^{3/4}$. It shows a positive dispersion, and remains visible in the spectrum up to a momentum $q_{\rm max} \approx k_{\rm F} ( \sqrt{1+2\hbar \omega_{\rm P}/E_{\rm F}} -1 )$, above which it enters the particle-hole continuum and becomes Landau-damped. 

Having determined the surface response functions for water and for the jellium electrons, we may compute the {quantum friction coefficient of water} according to Eq.~\eqref{qfriction}, for a range of Fermi energy and effective mass values. Result are shown in Fig. 3c. The friction coefficient 
$\lambda_{\rm Q}$ is given in the standard unit $\rm N \cdot s \cdot m^{-3}$. In terms of the slip length $b = \eta/\lambda$, with $\eta$ the fluid viscosity, $\lambda =10^5 ~\rm N\cdot s \cdot m^{-3}$ corresponds to a slip length $b = 10~\rm nm$ for water. 
Generally, the quantum contribution to friction is found to be non-negligible when the electronic system has excitations at low energy and high momentum. For instance, we find it to be very small {for water on semiconductor surfaces, which, for our purpose, can be described by a jellium model with low Fermi energy and effective mass, regardless of the nature (electron or hole) of the charge carriers. In such systems, electronic excitations are restricted to very small momenta, and} we expect the hydrodynamic friction to be dominated either by the classical roughness term \cite{Barrat1999,Falk2010} 
or by the optical phonon contribution (SI section 5.3). 

{On metal surfaces}, with high Fermi energy (1 - 10 eV) and effective mass close to unity, we find $\lambda_{\rm Q} \sim 10^{2}~\rm N \cdot s \cdot m^{-3}$, which is two orders of magnitude lower than typical hydrodynamic friction coefficients. {In this case, the quantum contribution to the friction is expected to be smaller than the roughness-based contribution. However, to our knowledge,} there are no experimental measurements of water slip length on atomically smooth metal surfaces. \rev{We note that our theory does not address reactive (typically non-close-packed) metal surfaces, on which chemical bonding with water occurs.}
The highest values of quantum friction coefficient in the $(m^*,E_{\rm F})$ parameter space are obtained in the region of low Fermi energy and high effective mass. In this region, the electronic surface plasmon mode is at low enough energy ($\sim 10 ~\rm meV$) and high enough momenta ($\sim 0.5 ~\rm \AA^{-1}$) to couple with the Debye peak of water: the resonant coupling between these two modes results in a particular friction enhancement (see figure S3). 

\vskip0.5cm
\noindent{\bf \large The odd water-carbon interface}\\ 
\rev{The classical contribution to the water-carbon friction has been extensively studied in the framework of MD simulations.
On flat graphite surfaces, friction coefficient values in the range $\lambda_{\rm Cl} \approx 10^4 - 10^5~\rm N \cdot s \cdot m^{-3}$ have been reported numerically, depending on the chosen force field~\cite{Paniagua-Guerra2020}. In carbon nanotubes, these values were found to be unaffected by wall curvature for tube radii larger than 10~nm~\cite{Thomas2008,Falk2010}. This strongly suggests that the radius-dependence of water slippage observed experimentally in carbon nanotubes with radii between 15 and 50~nm~\cite{Secchi2016} cannot be explained by the classical contribution to friction alone. The experiments in ref.~\cite{Secchi2016} then set an upper bound for the water-carbon classical friction at the lowest total friction coefficient measured in large radius nanotubes, that is $\lambda_{\rm Cl}^{\rm max} = 3 \times 10^3~\rm N \cdot s \cdot m^{-3}$. Hence, MD simulations likely overestimate the water-graphite friction coefficient by at least a factor of 3, which is typical in simulations of other water-solid systems~\cite{Bocquet2010}.}

Therefore, we must consider in detail the possible contribution of quantum friction in various water-carbon systems. We first study a single graphene sheet, whose surface response function can be calculated analytically (\cite{Wunsch2006} and SI section 6.1). The result is plotted in Fig. 4a, considering a doping level $E_F = 0.1~\rm  eV$. Graphene is found to have low energy excitations ($ \omega \lesssim 100~\rm meV$) only at very small momenta ($q \lesssim 0.05~\rm \AA^{-1}$). An intra-layer plasmon mode is present, but it displays a very steep square root dispersion at small momenta. The quantum contribution to the water friction coefficient, evaluated with Eq.~\eqref{qfriction}, is accordingly found to be very small, below $10^0 ~\rm N\cdot s \cdot m^{-3}$, whatever the doping level, much like in the case of semiconductors treated within the jellium model. \rev{We conclude that hydrodynamic friction on monolayer graphene is dominated by the classical contribution and should therefore be very small ($\lambda \leq \lambda_{\rm Cl}^{\rm max}$). This is in line with recent measurements of water slippage on monolayer graphene deposited on a silica substrate~\cite{Duan2018}, which yielded friction coefficients as small as $4.5\times 10^3~\rm N \cdot s \cdot m^{-3}$ (slip length $b = 200~\rm nm$).  }

The situation must be different, however, for water on multilayer graphite, which was found experimentally to exhibit much higher hydrodynamic friction: $\lambda \approx 2\times10^4 -10^5 ~\rm N \cdot s \cdot m^{-3}$ on flat graphite surfaces~\cite{Maali2008,Radha2016} and $\lambda \approx 3 \times 10^4 ~\rm N \cdot s \cdot m^{-3}$ in multiwall carbon nanotubes with large (50~nm) radius~\cite{Secchi2016}.
Indeed, in a staggered stack of graphene sheets (Fig. 4b), electrons acquire an extra degree of freedom compared to monolayer graphene, as they may tunnel between the sheets. In particular, the coupling between second nearest layers is associated with a bandwidth $4\gamma_2 = 40~\rm meV$ (Fig. 4b), resulting in a markedly different low energy excitation spectrum. In electron energy loss spectroscopy, graphite was found to exhibit a surface plasmon mode, polarised perpendicularly to the layers, at $\omega_{\rm P} =  50~\rm meV$ (at 300 K), with a very flat dispersion in the measured momentum range, which was up to $q_{\rm max} = 0.2~\rm \AA^{-1}$ at 300 K \cite{Portail1999} and $q_{\rm max} = 0.4~\rm \AA^{-1}$ at 600 K \cite{Laitenberger1996} (see Fig. 4a).
Based on our study of the jellium model, we expect this low-energy plasmon of graphite to strongly interact with the water Debye mode, resulting in an enhancement of quantum friction. This enhancement may be assessed by describing, to the lowest level of approximation, the plasmon contribution to the graphite surface response in terms of a Drude model, \rev{which is based on the semi-classical treatment of free electron dynamics~\cite{Pitarke2007}}: 
\begin{equation}
g_{\rm e} (q,\omega) = \frac{\omega_{\rm P}^2}{\omega_{\rm P}^2 - \omega^2 - 2i \gamma \omega}\theta(q_{\rm max}-q),
\end{equation}
with $\gamma \sim 25~\rm meV$ the surface plasmon width~\cite{Laitenberger1996}, and $\theta$ the Heaviside step function. 
Using this expression in Eq.~\eqref{qfriction}, we obtain a water quantum friction coefficient $\lambda_{\rm Q} = 0.4 \times 10^3 ~\rm N \cdot s \cdot m^{-3}$ for $q_{\max} = 0.2 ~\rm \AA^{-1}$ and $\lambda_{\rm Q} = 5 \times 10^3 ~\rm N \cdot s \cdot m^{-3}$ for $q_{\max} = 0.4 ~\rm \AA^{-1}$, which is indeed orders of magnitude larger than the expectation for graphene (Fig. 4c), and \rev{non-negligible compared to the upper bound for the classical contribution $\lambda_{\rm Cl}^{\rm max}$. Hence, the crude Drude model estimate suggests that the quantum contribution to friction may account for the difference in hydrodynamic slippage between monolayer graphene and multilayer graphite. }

\begin{figure}
\centering
\includegraphics{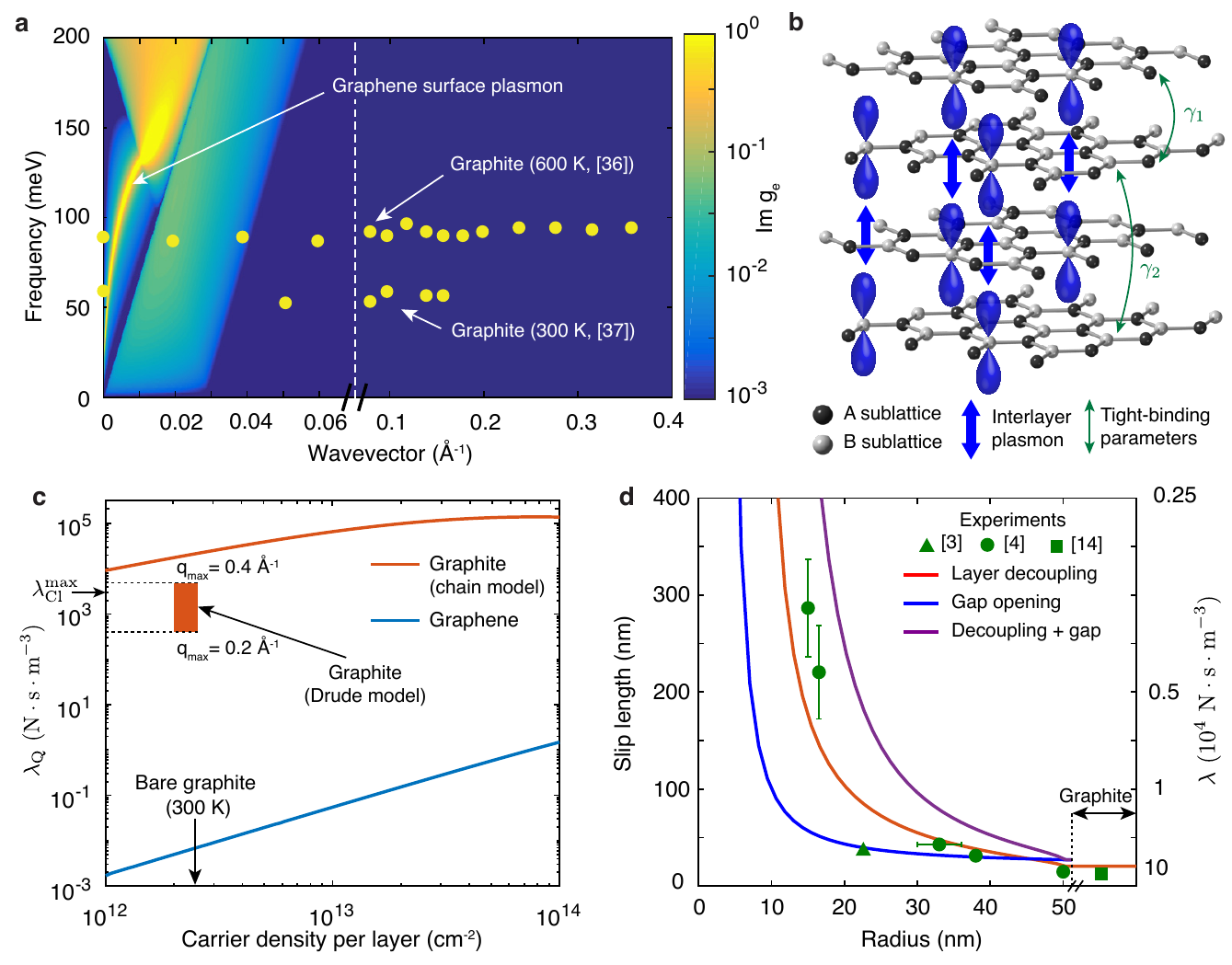}
\caption{\textbf{Quantum friction at the water-carbon interface}. \textbf{a}. Surface response function of doped graphene ($E_{\rm F} = 0.1 ~\rm eV$) in energy-momentum space, and experimental data~\cite{Laitenberger1996,Portail1999} for the graphite surface plasmon mode. \textbf{b}. Schematic of the electron movement corresponding to the graphite surface plasmon, and definition of the interlayer coupling parameters. \textbf{c}. Quantum friction coefficient for water on graphene and graphite as a function of charge carrier density. For graphene, the carrier density is determined by the doping level. For graphite, the friction coefficient is determined from the 1D chain model (Eq.~\eqref{1Dchain}), where the carrier density corresponds to the parameter $n_{\rm s}$. \textbf{d}. Water slip length ($b = \eta/\lambda$, with $\eta$ the water viscosity) in multiwall carbon nanotube as a function of inner tube radius. The green symbols are experimental data from refs. \cite{Secchi2016,Maali2008,Whitby2008}, and the full lines are theoretical predictions, corresponding to different models for the radius dependence of $n_{\rm s}$ in Eq.~\eqref{1Dchain}. }
\end{figure}

Going beyond a phenomenological treatment for the water-graphite quantum friction is particularly challenging, due to the very large unit cell required for numerically treating low energy and high momentum surface excitations~\cite{Lavor2020}. Nevertheless, we propose a simplified model in {order} to extract the physical ingredients at play in the graphite surface response. 
At a qualitative level, the free charge carriers that contribute to the low energy plasmon are located mainly on the $B$ sublattice~\cite{Tomanek1987} (Fig. 4b).
The flat plasmon dispersion has been attributed to the shape of the bands containing those free carriers~\cite{Laitenberger1996}. As a consequence of interlayer coupling, these are nearly flat up to parallel momentum $k \sim \gamma_1/v_{\rm F} = 0.06- 0.11~\rm \AA^{-1}$, with $\gamma_1$ the nearest-neighbour interlayer coupling parameter (Fig. 4b) and $v_{\rm F}$ the graphene Fermi velocity. If the electron dispersion is flat parallel to the layers, then the graphite can be pictured as an array of independent 1D chains extending perpendicular to the layers, at least within a certain momentum range. This assembly of localised oscillators are then expected to have excitations whose energy does not depend on wavevector, that is a dispersionless mode, {as highlighted experimentally.} 

Following this idea, we consider only the atoms on the $B$ sublattice, which form an assembly of tight-binding chains with coupling parameter $\gamma_2 = 10~\rm meV$~\cite{Partoens2006}. At sufficiently large momenta ($q \gtrsim 1/(2c) = 0.14~\rm \AA^{-1}$, with $c = 3.35~\rm \AA$ the interlayer spacing), we may consider that the external field acts only on the topmost (surface) atoms. We then compute the local density response of the topmost atom $a$ of a 1D chain $\delta n_a(q,\omega) = \chi_a(q,\omega) \phi_a(q,\omega)$, where $\phi_a$ is the potential acting on atom $a$ (see SI section 6.2). Then, treating the Coulomb interactions between the chains in the random phase approximation, we obtain the graphite surface response function as 
\begin{equation}
g_{\rm e} (q,\omega) = \frac{n_{\rm s} v_q \chi_a(q,\omega)}{n_{\rm s} v_q \chi_a(q,\omega)-1}, 
\label{1Dchain}
\end{equation}
where $v_q = \frac{e^2}{4\pi \epsilon_0} \frac{2\pi}{q}$ is the 2D Coulomb potential and $n_{\rm s}$ is the density of charge carriers contributing to the low energy mode. Our simple model accounts indeed for an excitation continuum around $\omega = 40~\rm meV$, whose intensity slowly decays with momentum (figure S5). 
An estimate for the parameter $n_{\rm s}$ is provided by the free charge carrier density in bulk graphite at 300~K ($n_{\rm s} = 2.3 \times 10^{12}~\rm cm^{-2}$, inferred from \cite{Gruneis2008}, see SI section 6.2): this yields a friction coefficient $\lambda_{\rm Q} = 1.8 \times 10^4~\rm N \cdot s \cdot m^{-3}$, slightly larger than the Drude model estimates, and within the range of experimentally measured water friction coefficients on graphite (Fig. 4d).
\rev{Ultimately,} \rev{the value of $n_{\rm s}$ will depend on the details of the electronic structure at the graphite surface, which undergoes renormalisation in the presence of water. }
\rev{An increase in the density of states at low energy and high momenta due to electron scattering on water fluctuations may lead to a higher apparent carrier density $n_{\rm s}$ (the upper limit being the total electron density on the $B$ sublattice, $4.3\times 10^{14}~\rm cm^{-2}$) and to a further increase in the expected value for the quantum friction coefficient (Fig. 4c).}

\rev{Overall, our theory predicts a strong difference in quantum friction between the water-graphene and water-graphite interfaces, explaining how water can exhibit larger slippage on monolayer graphene~\cite{Duan2018} than on multilayer graphite~\cite{Maali2008,Secchi2016,Radha2016}}. This counterintuitive difference further suggests an explanation for the radius-dependent water slippage in multiwall carbon nanotubes~\cite{Secchi2016}. In these quite large nanotubes ($15-50~\rm nm$ radius), \rev{while the water surface response is unaffected by confinement (SI section 4)}, the interlayer coupling is known to strongly depend on radius. A 50 nm radius tube has locally a graphite-like structure, while in a 10 nm radius tube the shells are completely decoupled~\cite{Endo1997}. Therefore, in large radius nanotubes, water is subject to graphite-like high quantum friction, while in smaller radius nanotubes water experiences graphene-like low friction, likely dominated by the classical contribution. 

The quantitative relevance of this argument can be checked, at the simplest level, in the framework of the Franklin model~\cite{Franklin1951,Speck1989}, which relates the probability $p$ of two layers being misaligned in a graphite structure to the average interlayer spacing $d$: $d = 3.44 - 0.086\cdot (1- p^2) ~ \rm (\AA)$. The dependence of $d$ on the inner radius $R$ for a multiwall nanotube can be inferred from experiment~\cite{Endo1997} (see SI section 6.4). We may then assume that the electron density $n_{\rm s}$ in Eq.~\eqref{1Dchain} scales according to $n_{\rm s}(R) = n_{\rm s}^0 (1-p(R))$, and we choose $n_{\rm s}^0 = 10^{13}~\rm cm^{-2}$ so that $\lambda (R \to \infty) = 4.4 \times 10^4 ~\rm N \cdot s \cdot m^{-3}$. The resulting prediction for the slip length is shown in Fig. 4d, and is found to be in good agreement with the experimental data. We note that if the inner shell tubes are semiconducting, the radius-dependent band gap $E_{\rm g} (R) = (2/3) v_{\rm F}/ R$~\cite{Charlier2007} may also reduce the number of charge carriers contributing to low energy excitations: we then expect a scaling $n_{\rm s} (R) = n_{\rm s}^0 e^{-E_{\rm g}(R)/k_{\rm B} T}$. Such a scaling also provides reasonable agreement with the data (Fig. 4d). While the details of electronic excitations in multiwall carbon nanotubes are hard to investigate theoretically, our theory strongly suggests that quantum friction is a key ingredient for determining the water slip length in these systems. 
    
\vskip0.5cm
\noindent{\bf \large Conclusions}\\
Our theory predicts quantum effects in the dynamics of the solid-liquid interface. We show that hydrodynamic friction results not only from the static roughness of the solid surface, but also from the coupling of water fluctuations to electronic excitations within the solid: this contribution is termed "quantum friction". Quantum friction is uniquely revealed at the water-carbon interface. We demonstrate that water friction is not anomalously low on graphene: it is rather anomalously high on graphite, due to the quantum contribution resulting from a graphite-specific terahertz plasmon mode. This finding enables us to rationalise the peculiar friction properties of water on carbon surfaces, and in particular the puzzling radius dependence of slippage in carbon nanotubes. 

More generally, quantum friction may be an important contribution to hydrodynamic friction on atomically smooth surfaces, provided that these have electronic excitations at low energy ($\omega \lesssim 100~\rm meV$) and high momenta ($q \sim 0.5 ~\rm \AA^{-1}$), which may couple to the Debye mode of water. \rev{However, true atomic smoothness cannot be achieved for any material, since many surfaces may become oxidised or charged in water. This is the case of insulating hexagonal boron nitride (hBN), which is expected to have negligible quantum hydrodynamic friction, so that water-hBN friction should be dominated by a small classical contribution as in the case of graphene. Yet, experimentally, water friction coefficients in hBN nanotubes are particularly high, corresponding to slip lengths $b < 5~\rm nm$~\cite{Secchi2016}. This high friction is attributed to the strong charging of hBN surfaces in water due to the chemisorption of OH$^-$ ions ~\cite{Siria2013}, which increases the surface roughness and results in a strong enhancement of classical friction, as has been shown in MD simulations~\cite{Xie2020a}.}

We stress that quantum friction is beyond the reach of standard \emph{ab initio} simulations~\cite{Tocci2014}, since it is an effect of electronic excitations beyond the Born-Oppenheimer (BO) approximation. At the water-carbon interface, the breakdown of the BO approximation is particularly strong: this is in line with the recent observation of non-BO effects on the graphene optical phonon frequencies~\cite{Pisana2007}. Therefore, new simulation methods will be required to tackle the dynamics of such interfaces, with the perspective of controlling nanoscale water flows thanks to the vastly tuneable electronic properties of carbon-based confining walls.

\vskip0.5cm
\noindent{\bf \large Acknowledgements}\\
The authors thank A. Robert for help with molecular dynamics simulations, and acknowledge fruitful discussions with B. Dou\c{c}ot, R. Netz, B. Coasne, N. Lorente, and B. Rotenberg. L.B. acknowledges funding from the EU H2020 Framework Programme/ERC Advanced
Grant agreement number 785911-Shadoks and ANR project Neptune. This work has
received the support of "Institut Pierre-Gilles de Gennes", program ANR-10-IDEX-0001-02
PSL and ANR-10-LABX-31. The authors acknowledge the French HPC resources of GENCI for the grant A9-A0070807364.

\vskip0.5cm
\noindent{\bf \large Data availability}\\
The data that support the findings of this study are either in the Supplementary Information, or available on Zenodo (doi:10.5281/zenodo.5242930). 

\vskip0.5cm
\noindent{\bf \large Author contributions}\\
L.B., M.-L.B. and N.K. conceived the project. N.K. developed the theoretical framework. N.K. and L.B. co-wrote the paper, with inputs from M.-L.B. All authors discussed the results and commented on the manuscript. 

\vskip0.5cm
\noindent{\bf \large Competing interests}\\
The authors declare no competing interests. 

\vskip0.5cm
\noindent{\bf \large Correspondence and requests for materials}\\
should be addressed to N.K. or L.B.

\end{document}


\begin{titlepage}

\vspace*{2cm} \begin{center}
\large Supplementary information for: \par\nobreak
\vspace{20pt}\hrule\vspace{10pt}
\huge\textbf{
Fluctuation-induced quantum friction\\ in nanoscale water flows} \par\nobreak
\vspace{10pt}\hrule\vspace{.6cm}
  \vspace{.5cm} 
\large N. Kavokine, M.-L. Bocquet and L. Bocquet  \\
\vspace{1cm}
-\\
\vspace{1cm}
\end{center}
\vspace{2cm}
\tableofcontents
\end{titlepage}

\emph{We use SI units throughout the text. We adopt the following convention for the $n$-dimensional Fourier transform:}
\begin{equation}
\tilde f (\q) = \int_{-\infty}^{+\infty} \d^n \rr\, f(\rr) e^{-i \q \cdot \rr} ~~~~ \mathrm{and} ~~~~ f(\rr) =  \frac{1}{(2\pi)^n}\int_{-\infty}^{+\infty} \d^n \q\, \tilde f(\q) e^{i \q \cdot \rr}.
\end{equation} 

\section{Electronic friction on a single particle}

We use cylindrical coordinates $\rr = (\brho,z)$. We consider a point charge $e$ at $\rr_0 = (0,h)$ above a solid occupying the half-space $z <0$. This charge generates at every point $\rr$ a Coulomb potential $V(\rr-\rr_0)$, with
\begin{equation}
V(\rr) = \frac{e^2}{4 \pi \epsilon_0 \lVert \rr \rVert }. 
\end{equation}
with $\epsilon_0$ the vacuum permittivity. If the charge moves at velocity $\vv$ parallel to the surface, the Coulomb potential becomes time-dependent: $V(\rr,t) = V(\rr-(\rr_0+\vv t))$. The solid responds to this external potential by creating a polarisation charge density $\delta n (\rr,t)$. In the framework of linear response theory, a linear relation is assumed between induced density and external potential:
\begin{equation}
\delta n (\rr,t) = \int_{-\infty}^{+\infty} \d t' \int \d \rr' \chi_{\rm e}(\rr,\rr',t-t') V(\rr',t'). 
\label{chidef}
\end{equation}
This defines the (retarded) density response function $\chi_{\rm e}$ of the solid. The polarisation charge generates a Coulomb potential that acts back on the external charge. The corresponding force reads 
\begin{equation}
\mathbf{f}(t)  = - \int_{-\infty}^{+\infty} \d t' \int \d \rr \d \rr'\,  \nabla_{\rr_0} V(\rr_0+ \vv t - \rr)  \chi_{\rm e}(\rr,\rr',t-t') V(\rr'-\rr_0 - \vv t'). 
\end{equation}
We assume that the solid is translationally invariant parallel to the surface, and consider only the in-plane component of the force. Then, $\chi_{\rm e}(\rr,\rr')$ depends only on $(\brho-\brho',z,z')$ and we may carry out Fourier transforms with respect to the in-plane coordinate: 
\begin{equation}
\mathbf{f}(t)  = -\int \frac{\d \q}{(2\pi)^2} \, (i \q) \int_{-\infty}^0 \d z \d z' V_q(z+h) V_q(z'+h) \int_{-\infty}^{+\infty} \d t' e^{i\q \vv (t-t')} \chi_{\rm e} (q,z,z',t-t'),
\end{equation}
with 
\begin{equation}
V_q(z) = \frac{e^2}{4 \pi \epsilon_0} \frac{2\pi}{q} e^{-q |z|}
\label{FTV}
\end{equation}
the 2D Fourier transform of the Coulomb potential. Identifying the Fourier transform of $\chi_{\rm e}$ with respect to time, we obtain
\begin{equation}
\mathbf{f} =  \frac{e^2}{8 \pi^2 \epsilon_0} \int \d \q \, \frac{i \q}{q} e^{-2 qh} \left[ \frac{-e^2}{2\epsilon_0 q} \int_{-\infty}^{0} \d z \d z' \, e^{q(z+z')} \chi_{\rm e}(q,z,z',\omega =  \q \vv) \right]. 
\label{onepart}
\end{equation}
The expression in brackets defines the solid's surface response function $g_{\rm e} (q,\omega =  \q \vv)$. Since $g_{\rm e} (\rr,t)$ is a real function, its space-time Fourier transform satisfies $g_{\rm e}(q,-\omega) = g_{\rm e}(q,\omega)^*$. Hence $\mathrm{Re} \, g_{\rm e}(q,-\q \vv)$ is even under $\q \to - \q$ and vanishes upon integration in eq.~\eqref{onepart}. Therefore, we obtain 
\begin{equation}
\mathbf{f} =  \frac{-e^2}{8 \pi^2 \epsilon_0} \int \d \q \, \frac{\q}{q} e^{-2 qh}\, \mathrm{Im} \, g_{\rm e} (q, \q \vv),
\end{equation}
which is equation (2) of the main text. A similar result was obtained in~\cite{Persson1995}, at it generalises the classical bulk result~\cite{Zwanzig1970,Sedlmeier2014} to interfacial friction. 

\section{Many-body theory of solid-liquid friction}

\subsection{Model definition}

We consider a semi-infinite solid extending into the half-space $z<0$ in contact with a semi-infinite liquid (water) extending into the half-space $z>0$, at temperature $T$. The whole system is described within a quantum field theory framework. For simplicity, we describe only the electronic degrees of freedom of the solid by the creation and annihilation  fields $\Psi^{\dagger}(\rr,t)$ and $\Psi (\rr,t)$, respectively; we note that one could follow the steps described below with the addition of lattice degrees of freedom as a phonon field. The liquid is described by its charge density $n_{\rm w} (\rr,t)$, assumed to have gaussian fluctuations fully determined by the two-point functions $\langle n_{\rm w} (\rr,t) n_{\rm w}(\rr',t')\rangle$, which are treated as inputs of the model. A flow parallel to the interface is induced in the liquid, and the system is assumed to have reached a non-equilibrium steady state. We assume that the liquid flow is slow enough so that it does not affect the form of microscopic correlations in the liquid. In that case, the effect of the flow field $\vv$ is to shift the coordinates within the liquid according to $n_{\rm w} (\rr,t) \mapsto n_{\rm w}(\rr - \vv t,t)$. 

The liquid and the solid interact through long range Coulomb forces, and through short-range forces due to the coupling of electronic degrees of freedom of the solid with electronic degrees of freedom of the liquid, which we do not treat explicitly. If the frontier orbitals of the solid and the liquid are sufficiently far apart in energy, then these short-range forces amount simply to "Pauli repulsion", which prevents the solid and the liquid from interpenetrating each other~\cite{Salem1961,Rackers2019}. If not, there may be chemisorption of the liquid on the solid~\cite{Muscat1978}, which is a situation beyond the scope of this work. The former assumption is justified in particular for the water-carbon interface, where \emph{ab initio} simulations show no water chemisorption~\cite{Grosjean2019}. In the following, we will treat explicitly only the long-range Coulomb forces, with the effect of Pauli repulsion taken into account in the bare correlation functions, computed for semi-infinite media. 

The Coulomb interactions in the system are described by the Hamiltonian
\begin{equation}
\hat H_{\rm int} = \int \d \rr \d \rr' \hat n_{\rm e}(\rr',t) V(\rr- \rr') \hat n_{\rm w} (\rr-\vv t,t) + \frac{1}{2} \int \d \rr \d \rr' \hat n_{\rm e}(\rr',t) V(\rr-\rr') \hat n_{\rm e}(\rr,t),
\label{hint}
\end{equation}
where $n_{\rm e}(\rr,t) \equiv \Psi^{\dagger}(\rr,t) \Psi(\rr,t)$ is the electron density: the first term is the water-electron, and the second term is the electron-electron Coulomb interaction. Within the interacting system, we wish to compute the solid-liquid friction force 
\begin{equation}
\langle \mathbf{ \hat  F}(t) \rangle = -\int \d \rr \d \rr' \, \nabla_{\rr'}V (\rr - \rr') \langle \hat n_{\rm w}(\rr'-\vv t,t) \hat n_{\rm e} (\rr,t)\rangle,
\label{force_micro}
\end{equation}
where the average is taken over all quantum and thermal fluctuations in the system. Therefore, computing the friction force amounts to computing the equal-time water-electron density correlation function. We proceed by treating $\hat H_{\rm int}$ as a perturbation, and expanding the average in eq.~\eqref{force_micro} in powers of $\hat H_{\rm int}$ to arbitrary order. Because we are dealing with a non-equilibrium steady state, we do so in the Schwinger-Keldysh framework of perturbation theory. 

\subsection{Brief overview of the Keldysh framework}

We make use of the out-of-equilibrium perturbation theory formalism originally proposed by L.~V. Keldysh in 1965~\cite{Keldysh1965}, which has since then been extensively described in several books~\cite{rammer_2007,kamenev_2011} and reviews~\cite{Rammer1986}. However, in order to keep this Supplementary Information as self-contained as possible, we give here a brief introduction, whose formulation is largely based on~\cite{kamenev_2011}. The reader who is familiar with the Keldysh formalism may directly skip to section 2.3. 

Our solid-liquid system is governed by the total Hamiltonian $\mathcal{H}(t) = \hat H_0 + \hat H_{\rm int} (t)$, where $\hat H_0$ is the quadratic Hamiltonian describing the system at equilibrium with Coulomb interactions switched off. Suppose we wish to compute the mean value of the Schr\"odinger picture operator $\mathcal{O}$ at time $t$. It is defined by 
\begin{equation}
\langle \mathcal{O} \rangle(t) = \frac{\mathrm{Tr} [\hat\rho(t) \mathcal{O}]}{\mathrm{Tr}[\hat\rho(t)]},
\label{meano}
\end{equation}
where $\hat\rho(t)$ is the density matrix of the out-of-equilibrium interacting system. In order to evaluate $\hat\rho(t)$, we assume that the Coulomb interactions and the fluid flow, which sets the system out of equilibrium, are adiabatically switched on starting at $t = - \infty$. Then, we may express the interacting density matrix at time $t$ as a function of the non-interacting density matrix $\hat\rho_0$ at $t = -\infty$, and the evolution operator $\mathcal{U}$: 
$\hat\rho(t) = \mathcal{U}_{t,-\infty}\hat \rho_0 \mathcal{U}_{-\infty,t}$,
with 
\begin{equation}
\mathcal{U}_{t,t'} = \mathbb{T} \left[ \exp\left(-\frac{i}{\hbar} \int_{t'}^t \mathcal{H}(t) \d t \right) \right],
\end{equation}
$\mathbb{T}$ being the time-ordering operator. The average value in eq.~\eqref{meano} then becomes 
\begin{equation}
\langle \mathcal{O} \rangle(t) = \frac{\mathrm{Tr} [\mathcal{U}_{-\infty,t} \mathcal{O} \mathcal{U}_{t,-\infty} \hat\rho_0]}{\mathrm{Tr}[\hat\rho_0]},
\end{equation}
where we have used circular permutation within the trace. The expression under the trace can be read (from left to right) in terms of time evolution: the system is evolved from $t = -\infty$ where the density matrix is known, to $t$ where the observable is computed, and the back to $t = -\infty$. In equilibrium perturbation theory, one further assumes that the interactions are adiabatically switched off at $t = + \infty$, so that the state of the system at $t = +\infty$ differs from the state at $t= -\infty$ only by a phase factor. Then, instead of evolving the system from $t$ to $-\infty$, one can evolve it from $t = + \infty$, thereby avoiding the complication of forward-backward time evolution. After a non-equilibrium evolution, however, the system has no reason to go back to its initial state, even if the interactions are switched off. Hence, one has to consider the evolution of the system on a contour that goes forwards then backwards in time. In practice, one defines the Schwinger-Keldysh closed time contour $c$, which travels from $t = -\infty$ to $t = +\infty$ and then back. It can then be shown that
\begin{equation}
\langle \mathcal{O} \rangle(t) = \frac{1}{\mathrm{Tr}[\hat\rho_0]} \mathrm{Tr} \left[ \hat\rho_0 \mathbb{T}_c \cdot \mathcal{O}^{H_0}(t) e^{-\frac{i}{\hbar} \int_c H_{\rm int}^{H_0}(t') \d t'} \right],
\label{otc}
\end{equation}
where the subscript $H_0$ indicates operators in the Heisenberg picture with respect to $H_0$, and $\mathbb{T}_c$ is the time-ordering operator along the contour $c$. 

In practice, instead of average values of operators at one point in time, we will be interested in computing contour-ordered two-point functions of operators taken at different points in time, in particular
\begin{equation}
\chi_{ ew} (e,w) \equiv  -\frac{i}{\hbar}\left \langle \mathbb{T}_c \left\{ n^{\cal H}_{\rm e}(\rr_{\rm e},t_{\rm e}) n^{\cal H}_{\rm w}(\rr_{\rm w},t_{\rm w}) \right\} \right \rangle. 
\end{equation} 
It can be shown that, in analogy with eq.~\eqref{otc}, 
\begin{equation}
\chi_{\rm ew}(e,w) = -\frac{i}{\hbar}\left \langle \mathbb{T}_c \left\{ n^{H_0}_{\rm e}(\rr_{\rm e},t_{\rm e}) n^{H_0}_{\rm w}(\rr_{\rm w},t_{\rm w})  e^{-\frac{i}{\hbar} \int_c H^{H_0}_{\rm int}(t') \d t'}\right\} \right \rangle_0,
\label{chitc}
\end{equation}
with $\langle \cdot \rangle_0 \equiv \mathrm{Tr}[\hat\rho_0 \cdot]/\mathrm{Tr}[\hat\rho_0]$. 

Under the form \eqref{chitc}, the correlation function $\chi_{\rm ew}$ can be evaluated as a perturbation series, by expanding the exponential to arbitrary order. Each term in the series consists in the average value of the contour-ordered product a certain number of field operators, taken with respect to the non-interacting density matrix $\hat \rho_0$. Since $\hat\rho_0$ is gaussian in the field operators, Wick's theorem applies, and those average values of many operators can be expressed as a convolution of two-operator correlation functions or Green's functions. 

However, the contour-ordered correlation function is a complicated object, since it has a different form depending on the part, forward ($c_1$) or backward ($c_2$), of the contour where its two time points are taken. It can actually be pictured as a $2\times 2$ matrix, whose entries contain the four possible choices. For example, 
\begin{equation}
\chi_{\rm ew} = \left( 
\begin{array}{cc}
\chi_{\rm ew}^{11} & \chi_{\rm ew}^{12} \\
\chi_{\rm ew}^{21} & \chi_{\rm ew}^{22} \\
\end{array}
\right),
\end{equation}
with $\chi_{\rm ew}^{ij}$ corresponding to $t_{\rm e} \in c_i$ and $t_{\rm w} \in c_j$. It can be shown that the perturbation theory is consistent with this matrix structure: the convolution of two matrix correlation functions corresponds to matrix multiplication, followed by space-time convolution of the resulting component pairings. However, the components $\chi^{ij}$ are not convenient quantities in terms of physical meaning. Therefore, it is customary to redefine the correlation function through a certain matrix transformation, for which different conventions exist. We shall adopt the trigonal representation of Larkin and Ovchinnikov~\cite{Larkin1975}, which is obtained through the transformation 
\begin{equation}
\chi \mapsto L \tau^3 \chi L^{\dagger} = \left( 
\begin{array}{cc}
\chi^R & \chi^K \\
0 & \chi^A \\
\end{array}
\right), 
\end{equation}
with the matrices 
\begin{equation}
L = \frac{1}{\sqrt{2}} \left(
\begin{array}{cc}
1 & -1 \\
1 & 1 
\end{array}
\right)
~~~\mathrm{and} ~~~ \tau^3 = \left(
\begin{array}{cc}
1 & 0 \\
0 & -1 
\end{array}
\right).
\end{equation}
This transformation reveals three physically meaningful components for $\chi$: the \emph{retarded}, \emph{advanced} and \emph{Keldysh} correlation functions. For the density cross-correlation function $\chi_{\rm ew}$, these components are defined as 
\begin{align}
&\chi^R_{\rm ew}(e,w) = - \frac{i}{\hbar} \theta(t_{\rm e}-t_{\rm w}) \langle [\hat n_{\rm e}(e), \hat n_{\rm w}(w) ] \rangle \\
\label{chir}
&\chi^A_{\rm ew}(e,w) = \frac{i}{\hbar} \theta(t_{\rm w}-t_{\rm e}) \langle [\hat n_{\rm e}(e), \hat n_{\rm w}(w) ] \rangle \\
&\chi^K_{\rm ew}(e,w) = - \frac{i}{\hbar}  \langle \{\hat n_{\rm e}(e), \hat n_{\rm w}(w) \} \rangle, 
\label{chik}
\end{align}
where $[,]$ is the commutator and $\{,\}$ the anticommutator; all the operators are in the Heisenberg picture with respect to $\hat H_0$, and we have used condensed notations of the type $e \equiv (\rr_{\rm e},t_{\rm e})$. Similar definitions hold for the components of the electron-electron and water-water correlation functions $\chi_{e}$ and $\chi_{w}$\footnote{Here we based our discussion on density-density correlation functions, because these are relevant for the solid-liquid problem. A more usual discussion in terms of Green's functions can be found, for example, in~\cite{Rammer1986}. }. In terms of these definitions, the solid-liquid friction force (eq.~\eqref{force_micro}) can be recast as
\begin{equation}
\begin{split}
\langle \mathbf{\hat F} \rangle &= -\int \d \rr_{\rm e} \d \rr_{\rm w}\,  \nabla_{r_{\rm w}} V(\rr_{\rm w}-\rr_{\rm e}) \cdot \frac{i\hbar}{2}\chi_{\rm ew}^K (\rr_{\rm e},t_{\rm e},\rr_{\rm w},0)|_{t_{\rm e}=0} \\
& = - \frac{i\hbar}{4\pi} \int_{-\infty}^{+\infty} \d \omega \int \d \rr_{\rm e} \d \rr_{\rm w}\,  \nabla_{r_{\rm w}} V(\rr_{\rm w}-\rr_{\rm e}) \chi_{\rm ew}^K (\rr_{\rm e},\rr_{\rm w},\omega). \\
\end{split}
\label{force}
\end{equation}

\subsection{Diagrammatic perturbation theory}

We have established in the previous section that evaluating the solid-liquid friction force amounts to computing the Keldysh component of the electron-water density correlation function, which supports a perturbative expansion in terms of the Coulomb interaction $H_{\rm int}$. We discuss the structure of this expansion using a Feynman diagram representation. We adopt the following notations for the propagators: 
\begin{equation}
\includegraphics{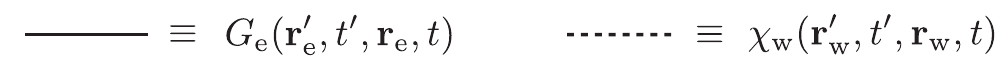}
\end{equation} 
Here $G_{\rm e}(e',e) = -(i/\hbar) \langle \mathbb{T}_c \cdot \Psi (e') \Psi^{\dagger}(e) \rangle_0$ is the bare electron Green's function. Recalling that $\hat n_{\rm e}(e) = \Psi^{\dagger}(e) \Psi(e)$, the Hamiltonian in eq.~\eqref{hint} allows for two types of vertices, corresponding to water-electron and electron-electron Coulomb interactions: 
\begin{equation}
\includegraphics{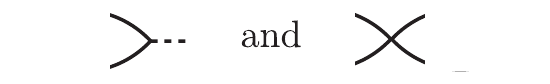}
\end{equation}
We start by considering only water-electron Coulomb interactions. Then, the series expansion of the exponential in eq.~\eqref{chitc} has the following diagrammatic representation: 
\begin{equation}
\includegraphics{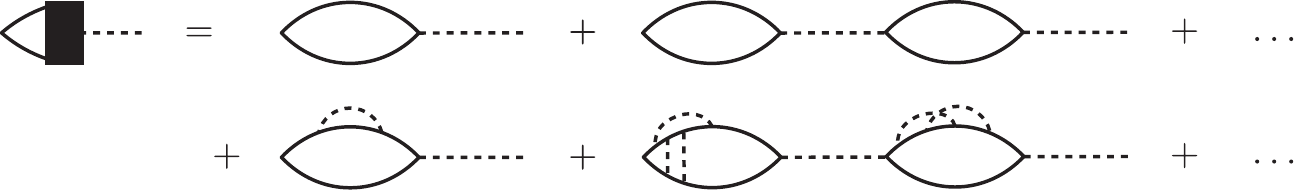}
\end{equation}
This expansion allows for partial resummation in the form of a Dyson equation:
\begin{equation}
\includegraphics{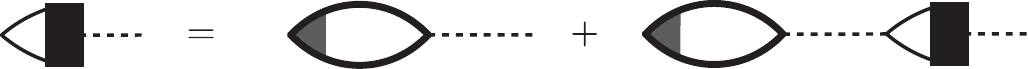}
\end{equation}
The thick line represents the electron Green's function renormalised by all self-energy corrections due to Coulomb interactions with water. The grey triangle represents vertex corrections, which are analogous to electron-phonon vertex corrections, and we expect them in general to be negligible according to the Migdal theorem~\cite{Migdal1958,Roy2014}. For simplicity, we drop these vertex corrections in the following, and leave their detailed investigation for future work. 

We now include the electron-electron Coulomb interactions at the self-consistent Hartree (RPA) level. This amounts to renormalise the electron polarisation bubble (or density correlation function) according to 
\begin{equation}
\includegraphics{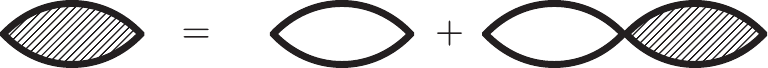}
\label{rpa}
\end{equation}
We note that in principle, electron-electron interactions beyond the RPA could be included, and would result in further self-energy and vertex corrections to the polarisation bubble. Ultimately, our Dyson equation for the electron-water density correlation function becomes
\begin{equation}
\includegraphics{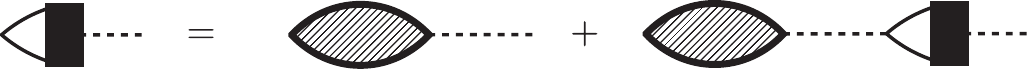}
\label{dyson}
\end{equation}

\subsection{General result}

We now explicit the analytical expressions corresponding to the Feynman diagrams. We denote $\chi_{\rm e}$ the renormalised electron polarisation bubble, as defined in eq.~\eqref{rpa}. Then, the first term in eq.~\eqref{dyson} reads 
\begin{align*}
\left(
\begin{array}{cc}
[\chi_{\rm ew}^{(1)}]^R(e,w) & [\chi_{\rm ew}^{(1)}]^K(e,w) \\
0 & [\chi_{\rm ew}^{(1)}]^A(e,w) 
\end{array}
\right) 
&\equiv [\chi_{\rm e} \otimes \chi_{\rm w}](e,w) \\
&=\int_{-\infty}^{+\infty} \d t \int  \d \rr_{\rm e}' \d \rr_{\rm w}' V(\rr_{\rm e}'-\rr_{\rm w}')\dots  \\
& \hspace{-3cm}  \left(
\begin{array}{cc}
\chi_{\rm e}^R(\rr_{\rm e},t_{\rm e},\rr_{\rm e}',t) & \chi_{\rm e}^K(\rr_{\rm e},t_{\rm e},\rr_{\rm e}',t) \\
0 & \chi_{\rm e}^A(\rr_{\rm e},t_{\rm e},\rr_{\rm e}',t) \\
\end{array}
\right)
\left(
\begin{array}{cc}
\chi_{\rm w}^R(\rr_{\rm w}'-\vv t,t,\rr_{\rm w},0) & \chi_{\rm w}^K(\rr_{\rm w}'-\vv t,t,\rr_{\rm w},0) \\
0 & \chi_{\rm w}^A(\rr_{\rm w}'-\vv t,t,\rr_{\rm w},0)  \\
\end{array}
\right).
\end{align*}
Having made explicit the definition of the convolution $\otimes$,  we may rewrite the Dyson equation~\eqref{dyson} in terms of the $R,A,K$ components: 
\begin{equation}
\left\{
\begin{array}{l}
\chi_{\rm ew}^K = \chi_{\rm e}^R \otimes \chi_{\rm w}^K + \chi_{\rm e}^K \otimes \chi_{\rm w}^A + \chi_{\rm e}^R \otimes \chi_{\rm w}^R \otimes \chi_{\rm ew}^K + (\chi_{\rm e}^R\otimes\chi_{\rm w}^K+\chi_{\rm e}^K\otimes \chi_{\rm w}^A)\otimes \chi_{\rm ew}^A\\
\\
\chi_{\rm ew}^{R,A} = \chi_{\rm e}^{R,A}\otimes \chi_{\rm w}^{R,A} + \chi_{\rm e}^{R,A} \otimes \chi_{\rm w}^{R,A} \otimes \chi_{\rm ew}^{R,A}
\end{array}
\right. .
\label{dyson2}
\end{equation}
Equation~\eqref{dyson2}, together with eq.~\eqref{force}, is our most general result, that holds far from equilibrium, and for any shape of the solid-liquid interface. 

\subsection{Classical contribution}
From now on, we will make several simplifying assumptions so as to obtain closed-form expressions for the solid-liquid friction coefficient. We start by splitting off the contribution to friction due to the solid's static roughness. The solid's charge density can always be split according to 
\begin{equation}
\hat n_{\rm e} (\rr,t) = \langle n_{\rm e}(\rr) \rangle + \delta \hat n_{\rm e}(\rr,t)  \equiv n_{\rm e}^0(\rr) + \delta \hat n_{\rm e}(\rr,t).
\end{equation}
The static charge density $n_{\rm e}^0(\rr)$ only contributes to the Keldysh component of the density correlation function $\chi_{\rm e}$. Precisely, 
\begin{equation} 
\begin{split}
\chi^K_{e}(\rr,t,\rr',t') &= - \frac{i}{\hbar} \left[ 2 n^0_{\rm e}(\rr)n^0_{\rm e}(\rr') + \langle \{\delta \hat n_{\rm e}(\rr,t), \delta \hat n_{\rm e}(\rr',t') \} \rangle \right] \\
&\equiv - \frac{i}{\hbar}  2 n^0_{\rm e}(\rr)n^0_{\rm e}(\rr') + \tilde \chi^K_{e}(\rr,t,\rr',t'). 
\end{split}
\label{chikstat}
\end{equation}
We consider the effect of static roughness only to first order in the solid-liquid Coulomb interaction. This would be exact in the case where the solid is completely static. If it is not, then we expect most electronic fluctuations to be at wavelengths that are large compared to the lattice spacing, so that the effect of these fluctuations on the surface roughness contribution is small. Nevertheless, the potential effect of these higher order contributions should be reassessed for any particular solid-liquid system under consideration. Using eqs. \eqref{force}, \eqref{dyson2} and \eqref{chikstat}, the first order surface roughness contribution to the friction force, which we call the \emph{classical} contribution, reads 
\begin{equation}
\mathbf{F}_{\rm Cl} = -  \int_{-\infty}^{+\infty} \d t \int \d \rr_{\rm e} \d \rr_{\rm e}' \d \rr_{\rm w} \d \rr_{\rm w}' \, \nabla_{r_{\rm w}}V(\rr_{\rm w}-\rr_{\rm e}) n_{\rm e}^0(\rr_{\rm e}) n_{\rm e}^0(\rr_{\rm e}') V(\rr_{\rm e}' - \rr_{\rm w}') \chi_{\rm w}^A(\rr_{\rm w}'-\vv t,t,\rr_{\rm w},0). 
\end{equation}
Since the liquid is translationally invariant parallel to the surface, we may carry out two-dimensional Fourier transforms. We obtain: 
\begin{equation}
\begin{split}
\mathbf{F}_{\rm Cl} = - \int_{-\infty}^{+\infty} \frac{\d \q}{(2\pi)^2} \, (i\q) &\int \d z_{\rm e} \d z_{\rm e}' \d z_{\rm w} \d z_{\rm w}' \, e^{-q|z_{\rm e}-z_{\rm w}|} e^{-q|z_{\rm e}'-z_{\rm w}'|}\dots \\
&\dots  V_q^2 n^0_{\rm e}(\q,z_{\rm e}) n^0_{\rm e}(-\q,z_{\rm e}') \int_{-\infty}^{+\infty} \d t e^{i \q \vv(z_{\rm w}') t} \chi_{\rm w}^A(q,z_{\rm w},z_{\rm w}',t), 
\end{split}
\end{equation}
with $V_q \equiv V_q(z = 0)$ (see eq.~\eqref{FTV}). We may now identify the Fourier transform of $\chi^A$ with respect to time. Then, we use that $\chi^A(q,z,z',-\omega) = \chi^A(q,z,z',\omega)^* = \chi^R(q,z,z',\omega)$ to obtain 
\begin{equation}
\begin{split}
\mathbf{F}_{\rm Cl} = \frac{e^2}{2\epsilon_0} \int\frac{\d \q}{(2\pi)^2} \, \frac{\q}{q} &\int \d z_{\rm e} \d z_{\rm e}' \d z_{\rm w} \d z_{\rm w}' \, e^{-q|z_{\rm e}-z_{\rm w}|} e^{-q|z_{\rm e}'-z_{\rm w}'|}\dots \\
&\dots  \frac{e^2}{2 \epsilon_0 q} n^0_{\rm e}(\q,z_{\rm e}) n^0_{\rm e}(-\q,z_{\rm e}')\mathrm{Im}[ \chi_{\rm w}^R(q,z_{\rm w},z_{\rm w}',\q\vv(z_{\rm w}'))].
\end{split}
\end{equation}
This result may be further simplified under the assumption that there is a well-defined separation between the solid and the liquid, say at $z = 0$, so that the integration over the $z_{\rm w}$'s runs over $[0, +\infty[$, while the integration over the $z_{\rm e}$'s runs over $]-\infty,0]$. As explained in the main text, we will further assume that the flow velocity $\vv$ is independent of $z$, which is true as long as the slip length (at least several nm on atomically smooth surfaces) is much larger than the range of the interactions contributing to the friction force (less than 1 nm). We then identify the (retarded) surface response function $g_{\rm w}(q,\omega)$ of the liquid, so as to obtain 
\begin{equation}
\mathbf{F}_{\rm Cl} = -\frac{e^2}{2\epsilon_0} \int \frac{\d \q}{(2\pi)^2} \, \frac{\q}{q} \int_{-\infty}^0 \d z_{\rm e} \d z_{\rm e}' e^{q(z_{\rm e}+z_{\rm e}')}n^0_{\rm e}(\q,z_{\rm e}) n^0_{\rm e}(-\q,z_{\rm e}') \, \mathrm{Im}[g^R_{\rm w}(q,\q\vv)]. 
\label{fcl1}
\end{equation}
This result may be cast into a more physically transparent form. First, we define 
\begin{equation}
|V_{\rm e}(\q)|^2 = \frac{1}{\mathcal{A}} \left|\frac{e^2}{2 \epsilon_0 q} \int_{-\infty}^0 \d z_{\rm e} \, e^{qz_{\rm e}} n_{\rm e}^0(\q,z_{\rm e}) \right|^2,
\end{equation}
which is the squared Fourier component at wavevector $\q$ of the average potential acting on the liquid at $z = 0$, normalised by the area $\mathcal{A}$ of the interface. Then, eq.~\eqref{fcl1} becomes 
\begin{equation}
\frac{\mathbf{F}_{\rm Cl}}{\mathcal{A}} = -\int \frac{\d \q}{(2\pi)^2} \, \q \, |V_{\rm e}(\q)|^2\left[ \frac{2 \epsilon_0 q}{e^2} \mathrm{Im}[g^R_{\rm w}(q,\q\vv)] \right]. 
\label{fcl2}
\end{equation}
Second, the expression in brackets can recast in terms of the liquid dynamic structure factor. We make use of the equilibrium relation between the Keldysh and retarded components of the surface correlation functions, which plays the role of a fluctuation-dissipation theorem~\cite{rammer_ch3}:
\begin{equation}
g^K_{\rm w}(q,\omega) = 2 i \, \mathrm{coth} \left( \frac{\hbar\omega}{2k_{\rm B}T} \right) \mathrm{Im} [ g_{\rm w}^R(q,\omega)]. 
\end{equation}
Taking the limit $\omega \to 0$ of this relation, and using the definition~\eqref{chik}, we obtain 
\begin{equation}
\lim_{\omega\to 0} \left(  \frac{2 \epsilon_0 q}{e^2} \frac{\mathrm{Im} [ g_{\rm w}^R(q,\omega)]}{\omega} \right)= \frac{1}{2\mathcal{A} k_B T} \int_{-\infty}^{+\infty} \d t \langle n_{\rm w}^s(\q,t) n_{\rm w}^s(-\q,0) \rangle,
\end{equation}
where 
\begin{equation}
n_{\rm w}^s(\q,t) = \int_0^{+\infty} \d z \, e^{-qz} n_{\rm w}(\q,z,t). 
\end{equation}
Defining the "surface" dynamic charge structure factor of the liquid as 
\begin{equation}
S_{\rm w}(q,t) = \frac{\langle n_{\rm w}^s(\q,t) n_{\rm w}^s(-\q,0) \rangle}{\mathcal{A}},
\end{equation}
and expanding eq.~\eqref{fcl2} to linear order in $\vv$, we obtain 
\begin{equation}
\frac{\mathbf{F}_{\rm Cl}}{\mathcal{A}} = - \frac{1}{8 \pi^2 k_B T}\int \d \q \, \q(\q\cdot \vv) \, |V_{\rm e}(\q)|^2\int_{-\infty}^{+\infty}\d t \, S_{\rm w}(q,t). 
\end{equation}
Finally, defining the classical friction coefficient $\lambda_{\rm cl}$ through $\mathbf{F_{\rm Cl}\cdot v}/\mathcal{A} = - \lambda_{\rm Cl} v^2$, we obtain 
\begin{equation}
\lambda_{\rm Cl} =  \frac{1}{4 \pi^2 k_B T}\int \d \q \, \frac{(\q\cdot \vv)^2}{v^2} \, |V_{\rm e}(\q)|^2\int_{0}^{+\infty}\d t \, S_{\rm w}(q,t),
\end{equation}
which is equation (5) of the main text. This is formally identical to the result obtained in the classical treatment of solid-liquid friction where the solid has no internal degrees of freedom~\cite{Barrat1999}; we recover it here in the corresponding limit of our quantum formalism. We note, however, that our formalism only deals with Coulomb interactions, and any roughness of the short-range Pauli repulsion forces is not taken into account. Nevertheless, this additional roughness could be dealt with in the exact same way as above, with the Coulomb potential replaced by a short-range repulsive potential, and the water charge density replaced by the water atom density. 

\subsection{Quantum contribution}

Having dealt with the static roughness contribution, we assume that the solid's average charge density vanishes: this amounts to replacing $\chi_{\rm e}^K \mapsto \tilde \chi_{\rm e}^K$ in eq.~\eqref{dyson2}. We then make the same simplifying assumptions as in the previous section: there is a separation between the solid and the liquid at $z=0$, and the velocity field does not depend on $z$. We further assume that the various correlation functions are translationally invariant parallel to the interface. This assumption is not necessary to proceed, but it greatly simplifies notations and is relevant for most practical purposes: we will be considering fluctuations at wavelengths that are longer than the solid's lattice spacing. 

With the above assumptions, we may define, for any correlation function $\chi$, the surface correlation function $g$ according to 
\begin{equation}
\begin{split}
g(\q,\omega) = \frac{-e^2}{4\pi \epsilon_0} \frac{2\pi}{q}  \int \d (\brho -\brho') &e^{-i \q (\brho-\brho')} \int_{-\infty}^{+\infty} \d (t-t') e^{i \omega (t-t') }\dots \\ & \dots \int_0^{+\infty} \d z \d z' e^{-q(|z|+|z'|)} 
\chi(\brho,\brho',z,z',t,t'). 
\end{split}
\end{equation}
These naturally satisfy the same Dyson equations~\eqref{dyson2} as the $\chi$ functions, and their convolution corresponds simply to multiplication in Fourier space. But, importantly, the effect of the flow velocty $\vv$ is to shift the frequencies appearing in water correlation functions according to $\omega \mapsto \omega - \q \vv$. With that, rearranging eq.~\eqref{dyson2} yields 
\begin{equation}
g_{\rm ew}^K(\q,\omega) = -\frac{g_{\rm e}^R(q,\omega) \, g_{\rm w}^K(q,\omega-\q \vv) + g_{\rm e}^K(q,\omega) \,g_{\rm w}^A(q,\omega-\q \vv)}{|1-g_{\rm e}^R(q,\omega)\,g_{\rm w}^R(q,\omega-\q \vv)|^2}, 
\label{gewK}
\end{equation}
where we have used that $g^R(\q,\omega) = [g^A(\q,\omega)]^*$. We now notice that Fourier-transforming eq.~\eqref{force} yields 
\begin{equation}
\frac{\langle \mathbf{\hat F} \rangle}{\mathcal{A}} = \frac{1}{2} \frac{i\hbar}{(2\pi)^3} \int_{-\infty}^{+\infty} \d \omega \int \d \q \, (i \q) g_{\rm ew}^K(\q,\omega). 
\label{Fg}
\end{equation}
In order to proceed, we make use of the fluctuation-dissipation theorem as in the previous section:
\begin{equation}
g_{e,w}^K(q,\omega) = 2i \, \mathrm{coth}\left( \frac{\hbar \omega}{2 k_B T} \right) \mathrm{Im} [g_{e,w}^R(q,\omega)] \equiv 2 i f(\omega) \mathrm{Im} [g_{e,w}^R(q,\omega)]. 
\end{equation}
We may then expand eq.~\eqref{gewK} as 
\begin{equation}
\begin{split}
g_{\rm ew}^K&(\q,\omega) =  2\frac{(f(\omega - \q \vv)-f(\omega) ) \mathrm{Im}[g_{\rm e}^R(q,\omega)] \, \mathrm{Im}[g_{\rm w}^R(q,\omega-\q \vv)] }{|1-g_{\rm e}^R(q,\omega)\,g_{\rm w}^R(q,\omega-\q \vv)|^2}\\
&- 2i\frac{f(\omega - \q \vv) \mathrm{Re}[g_{\rm e}^R(q,\omega)] \, \mathrm{Im}[g_{\rm w}^R(q,\omega-\q \vv)] +f(\omega ) \mathrm{Im}[g_{\rm e}^R(q,\omega)] \, \mathrm{Re}[g_{\rm w}^R(q,\omega-\q \vv)]}{|1-g_{\rm e}^R(q,\omega)\,g_{\rm w}^R(q,\omega-\q \vv)|^2}. 
\end{split}
\label{expansion}
\end{equation}
The imaginary parts of the surface response functions $g^R$ are odd functions of $\omega$, while the real parts are even. Therefore, the second term in eq.~\eqref{expansion} is even with respect to the transformation $\omega \mapsto - \omega, \q \mapsto -\q$ and vanishes upon integration in eq.~\eqref{Fg}. We hence obtain the quantum contribution to the friction force as:
\begin{equation}
\frac{\mathbf{F}_{\rm Q}}{\mathcal{A}} = \frac{\hbar}{(2\pi)^3}  \int \d \q \, \q \, \int_{-\infty}^{+\infty} \d \omega \, \frac{(f(\omega)-f(\omega-\q \vv) ) \mathrm{Im}[g_{\rm e}^R(q,\omega)] \, \mathrm{Im}[g_{\rm w}^R(q,\omega-\q \vv)] }{|1-g_{\rm e}^R(q,\omega)\,g_{\rm w}^R(q,\omega-\q \vv)|^2}.
\label{Fint}
\end{equation}
To linear order in $\vv$, the friction force becomes 
\begin{equation}
\frac{\mathbf{F}_{\rm Q}}{\mathcal{A}} = \frac{\hbar}{4 \pi^3}  \int \d \q \, \q \, (\q \cdot \vv) \,  \int_{0}^{+\infty} \d \omega \, \left( \frac{\d f (\omega)}{\d \omega} \right) \frac{ \mathrm{Im}[g_{\rm e}^R(q,\omega)] \, \mathrm{Im}[g_{\rm w}^R(q,\omega)] }{|1-g_{\rm e}^R(q,\omega)\,g_{\rm w}^R(q,\omega)|^2},
\end{equation}
with 
\begin{equation}
\frac{\d f(\omega)}{\d \omega} = -\frac{\hbar}{2k_B T} \, \mathrm{sinh}^{-2}\left( \frac{\hbar \omega}{2k_B T} \right). 
\end{equation}
Defining the quantum friction coefficient $\lambda_{\rm Q}$ by $\mathbf{F}_{\rm Q} /\mathcal{A} = - \lambda_{\rm Q} \vv$, we obtain after angular integration 
\begin{equation}
\lambda_{\rm Q} = \frac{\hbar^2}{8 \pi^2 k_B T} \int_0^{+\infty} \d q \, q^3  \int_{0}^{+\infty} \frac{\d \omega}{\mathrm{sinh}^{2}\left( \frac{\hbar \omega}{2k_B T} \right)} \frac{ \mathrm{Im}[g_{\rm e}^R(q,\omega)] \, \mathrm{Im}[g_{\rm w}^R(q,\omega)] }{|1-g_{\rm e}^R(q,\omega)\,g_{\rm w}^R(q,\omega)|^2},
\label{lambda}
\end{equation}
which is equation (6) of the main text after minor rearrangement. 

\subsection{Discussion}

Our result for the solid-liquid quantum friction coefficient (eq.~\eqref{lambda}) is formally identical to the one obtained in the solid-solid case by Volokitin and Persson (\cite{Volokitin2007} and references therein), and differs by a factor of 2 from the one obtained by Despoja, Echenique and Sunjic~\cite{Despoja2018}. However, none of the literature derivations for non-contact solid-solid friction may readily be extended to the solid-liquid case. Indeed, to our knowledge, the only derivation based on rigorous field theory arguments is restricted to the case of solids with local dielectric response~\cite{Volokitin2006}. At the solid-liquid interface, lengthscales as small as the solid's lattice spacing come into play, and an assumption of local dielectric response cannot hold. On the other hand, explicitly non-local derivations~\cite{Despoja2018,Persson1998} have treated the friction force rigorously to first order in the Coulomb interactions, and based the expression of higher order terms on an idea of multiple reflections of the electromagnetic field. This reasoning is difficult to justify fundamentally, especially when the two media are actually in contact, as in the solid-liquid case. 

Our formalism deals rigorously with the Coulomb interactions up to arbitrary order in fully non-local media. This allows us to obtain contributions beyond the first order (the denominator in eq.~\eqref{lambda}) without resorting to multiple reflection arguments, as the solution of a Dyson equation written in non-equilibrium perturbation theory. Beyond justifying the use of eq.~\eqref{lambda} for the solid-liquid interface, our formalism unambiguously specifies in which way -- if at all -- the response functions appearing in eq.~\eqref{lambda} should be renormalised by the solid-liquid interactions. Indeed, if interactions are taken into account at the RPA level, our computation shows that the response functions appearing in eq.~\eqref{lambda} are bare response functions. Therefore, they do not need further RPA renormalisation, for instance according to~\cite{Despoja2006}
\begin{equation}
\mathbf{g}^R_{e,w}(q,\omega) = \frac{g^R_{e,w}(q,\omega)}{1-g^R_{\rm e}(q,\omega)g^R_{\rm w}(q,\omega)}.
\label{gtot}
\end{equation}
The only interaction corrections that are not explicit in eq.~\eqref{lambda} are self-energy corrections to the electronic response function. 

\subsection{Toy model}

In this section, we present an elementary model that suggests a qualitative interpretation for our main result in eq.~\eqref{lambda}.
Let us assume that the charge fluctuations in each medium can be described by a single harmonic mode, with the same wavevector $\q$ and frequency $\omega_{\q}$. 
Then, the total hamiltonian of the system is 
\begin{equation}
\hat H = \hbar \omega_{\q} ( w^{\dagger}_{\q} w_{\q} + s^{\dagger}_{\q} s_{\q} ) + H_{\rm int}, ~~~~ H_{\rm int} = V_q (w^{\dagger}_{\q} s_{\q} + s^{\dagger}_{\q} w_{\q}).
\label{h2modes}
\end{equation}
where $V_q$ is a Fourier-transformed Coulomb interaction and $(w^{\dagger}_{\q},w_{\q})$ and $(s^{\dagger}_{\q},s_{\q})$ are the creation and annihilation  operators corresponding to the water and the solid modes, respectively. 
Elementary excitations (quasiparticles) may tunnel back and forth between the $s$ and $w$ modes. We denote $|n_{\rm s},n_{\rm w}\rangle$ the eigenstates of the non-interacting system. According to the Fermi golden rule, the quasiparticle tunnelling rate from mode $w$ to mode $s$ is 
\begin{equation}
\begin{split}
\gamma (w \to s) &= \frac{2\pi}{\hbar}\sum_{n_{\rm s},n_{\rm w}} P_{\rm s}(n_{\rm s}) P_{\rm w}(n_{\rm w}) | \langle n_{\rm w} -1, n_{\rm s}+1| H_{\rm int} | n_{\rm w},n_{\rm s} \rangle|^2 \\
&= \frac{2\pi}{\hbar}V_q^2 \sum_{n_{\rm s},n_{\rm w}} P_{\rm s}(n_{\rm s}) P_{\rm w}(n_{\rm w}) n_{\rm w} (n_{\rm s} + 1),
\end{split}
\end{equation}
where $P_{\rm s}(n_{\rm s})$ ($P_{\rm w}(n_{\rm w})$) is the probability that there are $n_{\rm s}$ ($n_{\rm w}$) quasiparticles in mode $s$ ($w$). Hence, the difference between forward and backward tunnelling rates is 
\begin{equation}
\Delta \gamma(\omega_{\q}) = \gamma(w\to s) - \gamma(s \to w) = \frac{2\pi}{\hbar} V_q^2 \sum_n n (P_{\rm w}(n)-P_{\rm s}(n)). 
\end{equation}
At equilibrium, 
\begin{equation}
\sum_n nP_{\rm w}(n) = \sum_n n P_{\rm s}(n) = n_{\rm B} (\omega_{\q}),
\end{equation}
with $n_{\rm B} (\omega) \equiv 1/(e^{\hbar \omega/k_B T} -1)$ the Bose distribution, so that the tunnelling rates compensate and there is no net momentum transfer between the two media. 
This is no longer true when the fluid flows with velocity $\vv$. Our full computation (section 2.6) shows that the flow induces
a Doppler shift in the occupation of the water modes. Precisely, the $w$ mode contains $n_{\rm B} (\omega_{\q}-\q\vv)$ quasiparticles, while the $s$ mode still contains $n_{\rm B}(\omega_{\q})$ quasiparticles. This results in a difference in between the $w\to s$ and $s\to w$ tunnelling rates:
\begin{equation}
\Delta \gamma (\omega_{\q}) = \frac{2\pi}{\hbar} V_q^2 (n_{\rm B} (\omega_{\q} - \q \vv) - n_{\rm B}(\omega_{\q})) 
 \approx  \pi V_q^2 \, \frac{\q\cdot \vv}{k_B T} \, \, \mathrm{sinh}^{-2} \left( \frac{\hbar \omega_{\q}}{2 k_B T} \right).
\label{deltag}
\end{equation}
Such an asymmetric quasiparticle tunnelling results in a net momentum transfer, 
hence an elementary friction force $\mathbf{f}_{\q} = \hbar \q \, \Delta \gamma (\omega_{\q})$. The total friction force is then obtained by integrating over all the modes: $\mathbf{F} / \mathcal{A} = \int \d^2 q \, \hbar \q \, \Delta \gamma(\omega_{\q})$. This is essentially the structure of eq.~\eqref{lambda}, where the frequency integral takes into account the distribution of modes at wavevector $\q$; the product of surface excitation spectra enforces energy conservation and contains the Coulomb interaction $V_q^2$. The mechanism of dissipation via asymmetric quasiparticle tunnelling between charge fluctuation modes is summarised in Fig. 1c of the main text. 

\section{Surface response function: general considerations} 

In our main formula for the quantum friction coefficient (eq.~\eqref{lambda} and eq. (6) of the main text), the properties of the solid and the liquid appear in the form of surface response functions. In this section, we give some general results concerning these objects. We recall the definition of the surface response function $g$, in terms of the density-density response function $\chi$ (see eqs.~\eqref{chidef} and \eqref{chir}) of a semi-infinite medium occupying the half-space $z<0$: 
\begin{equation}
g(q,\omega) = - \frac{e^2}{4\pi \epsilon_0} \frac{2\pi}{q} \int_{-\infty}^{0}\d z \d z' \,  e^{q(z+z')} \chi(q,z,z',\omega). 
\label{gdef}
\end{equation}

\subsection{Long wavelength limit}

It appears from the definition~\eqref{gdef} that the surface response function plays the role of a reflection coefficient for evanescent plane waves. Suppose the medium is subject to an evanescent plane wave at frequency $\omega$, of the form $\phi_{\rm ext}(\brho,z,\omega) = \phi_0 e^{i \q \brho} e^{q z}$. Its space-time Fourier transform is $\phi_{\rm ext}(z,q,\omega) = \phi_{0} e^{qz}$. Then, the induced charge density is 
 \begin{equation}
\delta n (q,z,\omega) = \phi_{0} \int_{-\infty}^0 \d z' \chi_q(z,z',\omega) e^{qz'},
\end{equation}
and the induced potential at a distance $z$ above the medium is 
\begin{equation}
\phi_{\rm ind}(q,z,\omega) = \phi_{0}  \int_{-\infty}^0 \d z' \d z'' \chi(q,z',z'',\omega) e^{qz'} \frac{e^2}{4 \pi \epsilon_0} \frac{2 \pi}{q} e^{-q(z-z")} = -g(q,\omega) \phi_{0} e^{-qz},
\end{equation}
or, in real space, $\phi_{\rm ind}(\rho,z,t) = - \phi_0 g(q,\omega) e^{i (\q \brho-\omega t)} e^{-q z}$. 

Given this interpretation, $g$ may be readily evaluated in the long wavelength limit, where the medium may be assumed to have a local dielectric response $\epsilon(\omega)$. In that case, inside the medium, there may be no induced charges (except on the surface), and the potential therefore solves the Laplace equation. Since it must vanish at $-\infty$, it is of the form $\phi_m(q,z) = \phi_m e^{qz}$. Outside the medium, the potential is the sum of the external potential and the induced potential. Since the Laplace equation holds, the outside potential reads 
\begin{equation}
\phi(q,z) = \phi_{\rm ext} e^{qz} + \phi_{\rm ind} e^{-qz}.
\end{equation}
Now, we must enforce boundary conditions on the interface. These are given by continuity of the potential and of the displacement field $\mathbf{D} = -\epsilon_0 \epsilon(\omega) \nabla \phi$. The boundary conditions read
\begin{equation}
\begin{split}
&\phi_{\rm ext} + \phi_{\rm ind} = \phi_m \\
&\phi_{\rm ext} - \phi_{\rm ind} = \epsilon (\omega) \phi_m.
\end{split}
\end{equation}
Hence we obtain $\phi_{\rm ind} = \phi_{\rm ext}(1-\epsilon(\omega))/(\epsilon(\omega)+1)$, and therefore the expression of $g$ in the long wavelength limit: 
\begin{equation}
g(q \to 0,\omega) = \frac{\epsilon(\omega)-1}{\epsilon(\omega) +1}. 
\end{equation}

\subsection{Specular reflection}

In general, computing the surface response requires the knowledge of the density response function $\chi(q,z,z',\omega)$ for a semi-infinite medium. For a liquid, it can be determined directly from classical molecular dynamics (MD) simulations (see section 4). For an electronic system, it can be computed within different analytical or numerical frameworks with varying degrees of accuracy. The simplest treatment (self-consistent Hartree, or RPA) that takes into account electron-electron interactions requires to solve the Dyson equation~\eqref{rpa}, which is written out explicitly as 
\begin{equation}
\chi(q,z,z',\omega) = \chi^0(q,z,z',\omega) + \int \d z_1 \d z_2 \, \chi^0(q,z,z_1,\omega) V_q(z_1-z_2) \chi (q,z_2,z',\omega). 
\label{dysonchi}
\end{equation}
Here $\chi^0$ is the non-interacting density response function. It can in principle be computed if the eigenenergies $E_{\lambda}$ and eigenfunctions $\psi_\lambda (\rr)$ of the non-interacting system are known~\cite{rammer_ch6}: 
\begin{equation}
\chi^0(\rr,\rr',\omega) = \sum_{\lambda,\lambda'}\, \frac{n_{\rm F}(E_{\lambda})-n_{\rm F}(E_{\lambda'})}{E_{\lambda}-E_{\lambda'} + \hbar \omega + i \delta}\, \psi^*_{\lambda}(\rr) \psi_{\lambda'}(\rr) \psi^*_{\lambda'}(\rr') \psi_{\lambda}(\rr'), 
\label{chi0def}
\end{equation}
where $n_{\rm F}$ is the Fermi-Dirac distribution and $\delta \to 0^+$. Even if $\chi^0$ is known, eq.~\eqref{dysonchi} must be solved numerically for every value of $q$ and $\omega$. A considerable simplification is achieved within the so-called specular reflection (SR) approximation, which allows one to solve~\eqref{dysonchi} analytically and express the surface response in terms of the bulk response. The SR approximation sets
\begin{equation}
\chi^0(q,z,z',\omega) = \chi^0_{\rm B} (q,z-z',\omega) + \chi^0_{\rm B} (q,z+z',\omega),
\label{specular}
\end{equation}
where $\chi^0_{\rm B}$ is the bulk system's non-interacting density response. This ansatz does not correspond to any particular form of the wavefunctions in eq.~\eqref{chi0def}. It imposes phenomenologically that in the presence of a surface, the points $z$ and $z'$ may either interact directly, or through a specular reflection from the surface at $z=0$. It can be shown that the SR approximation thus amounts to neglecting quantum interference between electrons impinging on and electrons reflected from the surface~\cite{Griffin1976}. 

Inserting eq.~\eqref{specular} into eq.~\eqref{dysonchi} and carrying out Fourier transforms along the vertical direction (the computation is detailed, for example, in \cite{Griffin1976}), one obtains for the surface response function the result reported in the main text: 
\begin{equation}
g (q,\omega) = \frac{1-q\ell_q(\omega)}{1+ q\ell_q(\omega)}, ~~~ \ell_q(\omega) = \frac{2}{\pi} \int_0^{+\infty} \frac{\d q_z}{(q^2+q_z^2)\epsilon(q,q_z,\omega)},
\label{gspecular}
\end{equation}
where $\epsilon(q,q_z,\omega) = 1 - \frac{e^2}{\epsilon_0 (q^2+q_z^2)} \chi^0_{\rm B}(q,q_z,\omega)$ is the bulk system's dielectric function. The bulk non-interacting density response function is obtained from the Fourier-transformed eq.~\eqref{chi0def}:
\begin{equation}
\chi^0_{\rm B} (q,q_z,\omega) =  \sum_{\nu,\nu'} \int_{\rm BZ} \frac{\mathrm{d}^3k}{4\pi^3} |\langle \mathbf{k} + \mathbf{q},\nu | e^{i \mathbf{q \cdot r}}| \mathbf{k} ,\nu'\rangle|^2\frac{n_{\rm F}[E_{\nu}(\mathbf{k+q})]-n_{\rm F}[E_{\nu'}(\mathbf{k})]}{E_{\nu}(\mathbf{k}+\mathbf{q})-E_{\nu'}(\mathbf{k})-\hbar (\omega +i\delta) },
\label{chi0bulk}
\end{equation}
where we have re-labeled the states $\lambda \mapsto (\mathbf{k},\nu)$, with $\nu$ a band index and $\mathbf{k}$ a vector within the (three-dimensional) first Brillouin zone. 

For completeness, we provide here an additional derivation of eq.~\eqref{gspecular}, which has the advantage of being computationally simpler than the one reported in \cite{Griffin1976}. It is based on the work of Ritchie and Marusak~\cite{Ritchie1966}, who first proposed the SR approximation in their study of surface plasmons.  The idea is that, when eq.~\eqref{specular} is enforced, the \emph{shape} of the density response of the semi-infinite medium to the potential $\phi_{\rm ext}(q,z,\omega) = \phi_{\rm ext}e^{qz}$ is the same as the \emph{shape} of the density response of an infinite medium to a symmetrised potential $\phi_{\rm eff} (q,z,\omega) = \phi_{\rm eff} e^{-q|z|}$. The amplitude $\phi_{\rm eff}$ is a priori non known, and it is determined by enforcing Maxwell boundary conditions at the interface. 

In the following, we will drop the frequency $\omega$ which plays no role in the computation. In response to the potential $\phi_{\rm eff}$, the induced charge density in the infinite medium reads
\begin{align}
\delta n (q,z) &= \phi_{\rm eff} \int_{-\infty}^{+\infty} \d z' \chi_{\rm B}(q,z-z') e^{-q|z'|}\\
& = \phi_{\rm eff}\frac{1}{2\pi} \int_{-\infty}^{+\infty} \d q_z \chi_{\rm B}(q,q_z) e^{iq_z z} \int_{-\infty}^{+\infty} \d z' e^{-q|z'|} e^{-i q_z z'} \\
& = \phi_{\rm eff} \frac{q}{\pi} \int_{-\infty}^{+\infty} \d q_z \frac{\chi_{\rm B}(q,q_z)}{q^2+q_z^2} e^{i q_z z}.
\end{align}
The induced potential $\phi_{\rm ind,m}$ (not to be confused with the induced potential $\phi_{\rm ind}e^{-qz}$ outside the medium) is 
\begin{align}
\phi_{\rm ind,m} (q,z)& = \int_{-\infty}^{+\infty} \d z' \, \delta n (q,z') \frac{e^2}{4\pi \epsilon_0} \frac{2\pi}{q} e^{-q|z-z'|} \\
& = 2 \phi_{\rm eff} \frac{e^2}{4 \pi \epsilon_0} \int_{-\infty}^{+\infty} \d q_z \frac{\chi_{\rm B}(q,q_z)}{q^2+q_z^2} e^{i q_z z} \int_{-\infty}^{+\infty} \d z' e^{-q |z-z'|} e^{i q_z z'} \\
& = 4 \phi_{\rm eff} \frac{e^2}{4\pi \epsilon_0} \int_{-\infty}^{+\infty} \d q_z \frac{q \chi_{\rm B}(q,q_z)}{(q^2+q_z^2)^2} e^{i q_z z}. 
\label{phiind}
\end{align}
At this point, we may introduce the bulk dielectric function $\epsilon(q,q_z)$. For the bulk interacting density response function, the RPA Dyson equation~\eqref{dysonchi} reduces to 
\begin{equation}
\chi_{\rm B}(q,q_z) = \frac{\chi^0_{\rm B}(q,q_z)}{1-\frac{e^2}{\epsilon_0 (q^2+q_z^2)}\chi^0_{\rm B} (q,q_z)}. 
\end{equation}
The dielectric function being defined according to $\epsilon(q,q_z) = 1 - \frac{e^2}{\epsilon_0 (q^2+q_z^2)} \chi^0_{\rm B}(q,q_z)$, we have the relation
\begin{equation}
\chi_{\rm B}(q,q_z) = \frac{\epsilon_0 (q_z^2+q^2)}{e^2} \left( \frac{1}{\epsilon(q,q_z)}-1 \right). 
\end{equation}
When inserting this relation into eq.~\eqref{phiind}, we need to compute the integral 
\begin{equation}
I(q) = \int_{-\infty}^{+\infty} \d q_z\, \frac{e^{i q_z z}}{q^2+q_z^2} = \frac{1}{q} \int_{-\infty}^{+\infty} \d u \, \frac{e^{iu qz}}{1+ u^2}.
\end{equation}
Specialising to the case $z<0$, and noticing that the integrand has poles at $i$ and $-i$, we may close the integration path in the lower complex plane, so that 
\begin{equation}
I(q) = - \frac{2 i \pi}{q} \underset{u= -i}{\mathrm{Res}} \left[\frac{e^{iu qz}}{1+ u^2} \right] = \frac{\pi}{q} e^{qz}. 
\end{equation}
Finally, 
\begin{equation}
\phi_{\rm ind,m} (q,z) = \phi_{\rm eff} \left(\frac{q}{\pi} \int_{-\infty}^{+\infty} \d q_z \,\frac{e^{i q_z z}}{(q^2+q_z^2) \epsilon(q,q_z)} -e^{qz} \right), 
\end{equation}
so that the total potential in the half-space $z<0$ is 
\begin{equation}
\phi_m(q,z) = \phi_{\rm eff} e^{qz} + \phi_{\rm ind,m}(q,z) = \phi_{\rm eff} \, \frac{q}{\pi} \int_{-\infty}^{+\infty} \d q_z \, \frac{e^{i q_z z}}{(q^2+q_z^2) \epsilon(q,q_z)}. 
\end{equation}
We now need to determine $\phi_{\rm eff}$ in the actual semi-infinite medium by enforcing the boundary conditions at the surface, which are, as in the local case (section 3.1), continuity of the potential and of the displacement field. Outside the medium, we may still express the potential as $\phi_{\rm ext} e^{qz} + \phi_{\rm ind} e^{-qz}$: the sum of the actual potential we are applying and the potential induced by the medium. The displacement field is produced only by the external charges, hence $\mathbf{D}(q,z) = - \epsilon_0 \nabla \phi_{\rm eff} (q,z)$ in the half-space $z<0$, so that the boundary conditions read:
\begin{equation}
\begin{split}
&\phi_{\rm ext} + \phi_{\rm ind} = q \ell_q \phi_{\rm eff} \\
&\phi_{\rm ext} - \phi_{\rm ind} =  \phi_{\rm eff}.
\end{split}
\end{equation}
We deduce
\begin{equation}
\phi_{\rm ind} = \frac{q \ell_q-1}{q \ell_q +1} \phi_{\rm ext}, 
\end{equation}
and, given that $\phi_{\rm ind} = - g(q,\omega) \phi_{\rm ext}$ (see section 3.1), we recover eq.~\eqref{gspecular}. 

\section{Surface response of water: molecular dynamics simulations}

We carried out classical molecular dynamics (MD) simulations of water in contact with a hydrophobic surface in order to determine the water surface response function. As a consistency check, we also carried out bulk water simulations, from which we extracted the frequency-dependent dielectric constant. \rev{Finally, we performed simulations of water slabs between two graphene sheets in order to assess the effect of confinement on the water surface response.} The complete simulation results are available on Zenodo~\cite{zenodo}. 

\subsection{Details of the simulations}

All simulations were carried out using the LAMMPS software~\cite{Plimpton1995}. We used the SPC/E water model~\cite{Berendsen1987} with the SHAKE algorithm~\cite{Ryckaert1977}. The simulations were carried out in the canonical (NVT) ensemble, with a stochastic CSVR thermostat~\cite{Bussi2007} with time constant 1~ps maintaining a temperature $T = 298.15~\rm K$. We used a timestep of 2~fs, and atomic positions were written out every 4~fs. Electrostatic interactions were calculated with a particle-mesh Ewald summation with a Coulomb cutoff of 1.4~nm.

\subsubsection{Bulk simulation}

The bulk simulation used $N = 8000$ water molecules. The volume of the simulation box was first adjusted in the NPT ensemble to yield a mass density $\rho = 0.99715~\rm g\cdot cm^{-3}$. The resulting volume was $V = (64.145)^3~\rm \AA^{3}$. The simulation was then equilibrated in the NVT ensemble for 200~ps, and the subsequent 20~ns were used for analysis. 

\subsubsection{Interface simulation}

The interface simulation was carried out with $N = 20200$ water molecules. The solid surface consisted of three graphene layers (with ABA stacking), with surface area $128.316 \times 123.490 ~\rm \AA^{2}$, and the simulation box height was 6.5~nm. The positions of the carbon atoms were frozen during the simulation. The direction normal to the surface was aperiodic, and spurious slab-slab interactions were removed. A reflective wall was placed close to the top edge of the box to prevent gaseous water molecules from crossing the top boundary. We used two sets of Lennard-Jones parameters for the water-carbon interaction, to which we refer to as "Werder" and "Aluru", listed in the following table. 
\begin{center}
\begin{tabular}{|c|c|c|c|c|c|}
\hline
Ref. & Name & $\epsilon_{\rm CO} ~\rm (kcal/mol)$ & $\sigma_{\rm CO} ~\rm (\AA)$ & $\epsilon_{\rm CH} ~\rm (kcal/mol)$ & $\sigma_{\rm CH} ~\rm (\AA)$ \\
\hline
\cite{Werder2003} & Werder & 0.0937 & 3.19 & - & - \\
\hline
\cite{Wu2013} & Aluru & 0.0850 & 3.436 & 0.0383 & 2.69 \\
\hline 
\end{tabular}
\end{center}
The simulation was equilibrated in the NVT ensemble for 200~ps and the subsequent 6~ns were used for analysis. 

\rev{\subsubsection{Slab simulations}

The slab simulations contained $N = 800, 1200, 1600$ or 2000 water molecules depending on the slab thickness. The two solid surfaces each consisted of two staggered graphene layers, with surface area $64.158 \times 61.745 ~\rm \AA^{2}$. The positions of the carbon atoms were frozen relative to each other. The distance between the surfaces was first equilibrated during 400~ps, then it was fixed, and the subsequent 6~ns were used for analysis. For each value of the number $N$ of water molecules, the following slab thickness was obtained: 
\begin{center}
\begin{tabular}{|c|c|}
\hline
$N$ & Slab thickness $h$ (\AA) \\
\hline
800 & 9.6 \\
1200 & 12.8 \\
1600 & 15.8 \\
2000 & 18.9 \\
\hline
\end{tabular}
\end{center}
"Werder" parameters were used for the water-carbon interaction. }

\subsection{Analysis of the simulations} 

The computation of response functions from equilibrium MD simulations is based on the fluctuation-dissipation theorem. 

\subsubsection{Bulk simulation}

The simulation was split into $N_{\rm s} = 20$ pieces of length $\Delta t = 1~\rm ns$, and the results obtained from each of the pieces were averaged to obtain the final result. The accessible frequencies were thus from $1~\rm GHz$ to $62.5~\rm THz$. At every sampled time $t$, we computed the Fourier-transformed water charge density $n_{\rm w}(\k,t) = \sum_i Z_i e^{i \k \rr_i(t)}$, with the index $i$ running over all the charged sites of the SPC/E water molecules, and $Z_i$ the corresponding charge. We define the dynamic charge structure factor according to 
\begin{equation}
S(\mathbf{k},\omega)= \frac{1}{\cal V}\int_{-\infty}^{+\infty}\text{d}t\langle n_{\rm w}(\mathbf{k},t)n_{\rm w}(-\mathbf{k},0)\rangle e^{i\omega t},
\label{structw}
\end{equation}
where $\cal V$ is the volume of the simulation box. 
Then, the fluctuation-dissipation theorem yields the susceptibility $\bar \chi$ according to: 
\begin{equation}
\mathrm{Im} \, \bar \chi( \k,\omega) = \frac{e^2}{ 2\epsilon_0 k^2} \frac{  \omega}{k_{\rm B}T}   S(\k,\omega).
\end{equation}
The susceptibility is related to the dielectric permittivity according to $\bar \chi(\k,\omega) = 1-1/\epsilon(\k,\omega)$. We therefore require also the real part of the susceptibility, which can be determined through a Kramers-Kr\"onig relation: 
\begin{equation}
\mathrm{Re} \, \bar \chi (\k,\omega) = \frac{2}{\pi} \mathcal{P} \int_0^{+\infty} \d \omega' \frac{\omega' \mathrm{Im}\, \bar \chi(\k,\omega')}{\omega'^2-\omega^2}, 
\end{equation}
where $\mathcal{P}$ indicates that the principal part of the integral is taken. In practice, the structure factor in eq.~\eqref{structw} was computed from the simulation data by making use of the Wiener-Khinchin theorem. The resulting spectra were convoluted with a gaussian filter of half-width $50~\rm GHz$. This allowed for some smoothing of the spectra, while not affecting their low-energy region, since the spectra are constant below 200~GHz. Then, spherical averaging was performed over the quantity $S(\k,\omega)/k^2$.

\subsubsection{Interface simulation}

The simulation was split into $N_{\rm s} = 60$ pieces of length $\Delta t = 0.1~\rm ns$, and the results obtained from each of the pieces were averaged to obtain the final result. The accessible frequencies were thus from $0.1~\rm GHz$ to $62.5~\rm THz$. At every sampled time $t$, we computed the Fourier-Laplace transform of the water charge density $n^s_{\rm w}(\q,t) = \sum_i Z_i e^{i \q \rr_i(t)}e^{-q z_i}$, with the index $i$ running over all the charged sites of the SPC/E water molecules, and $Z_i$ the corresponding charge. We define the dynamic surface structure factor according to 
\begin{equation}
S_{\rm s}(\mathbf{q},\omega)= \frac{1}{\cal A}\int_{-\infty}^{+\infty}\text{d}t\langle \delta n^s_{\rm w}(\q,t) \delta n^s_{\rm w}(-\mathbf{q},0)\rangle e^{i\omega t},
\label{structs}
\end{equation}
with $\cal A$ the surface area, and $\delta n_{\rm w}^s = n_{\rm w}^s - \langle n_{\rm w}^s \rangle$. 
We then obtain the imaginary part of the surface response function through the fluctuation-dissipation theorem:
\begin{equation}
\mathrm{Im} \, g_{\rm w}(\q,\omega) =  \frac{e^2}{4 \epsilon_0 q}  \frac{ \omega}{k_{\rm B}T}  S_{\rm s} (\q,\omega),
\label{FDT}
\end{equation}
In practice, the structure factor in eq.~\eqref{structs} was computed from the simulation data by making use of the Wiener-Khinchin theorem. The resulting spectra were convoluted with a gaussian filter of half-width $50~\rm GHz$. Then, radial averaging was performed over the quantity $S(\q,\omega)/q$. 

The surface response function evaluated in this way depends on the choice of origin for the vertical coordinate $z$. By default, the origin is placed at the first graphene plane, but this introduces a spurious vacuum gap (we call its thickness $d$) in the computation of the surface response. In fact, the procedure described above yields $e^{-2qd} g_{\rm w}(q,\omega)$ instead of $g_{\rm w}(q,\omega)$. 
However, there is no clear way of determining $d$ from microscopic considerations. Instead, we fixed $d$ by enforcing the compressibility sum rule for the surface response function in the long wavelength limit. The procedure is derived in detail in the Appendix. In brief, we first compute the static surface structure factor: $\bar S_{\rm s}(q) = (1/\mathcal{A}) \langle \delta n_{\rm w}^s(\q) \delta n_{\rm w}^s(-\q) \rangle$, with the average performed over the whole length of the simulation. Then, the compressibility sum rule reads
\begin{equation}
\frac{\pi e^2}{4 \epsilon_0 k_{\rm B} T} \frac{\bar S_{\rm s}(q)}{q} = \int_0^{\infty} \d \omega \, \frac{ \mathrm{Im} \, g_{\rm w}(q,\omega)}{\omega} = \frac{\pi}{2}g_{\rm w}(q,0) .
\label{compress}
\end{equation}
In the same way as the bulk structure factor has only even powers of $q$ in its low $q$ expansion, the surface structure factor is expanded only in odd powers of $q$ (see Appendix). Therefore $\bar S_{\rm s}(q)/q$ has a horizontal asymptote as $q \to 0$, while $e^{-2qd}\bar S_{\rm s}(q)/q$ has a linear scaling. Hence, in order to cancel the gap $d$, we adjust the origin of the coordinate $z$ so that $\bar S_{\rm s}(q)/q$ has indeed a horizontal asymptote. Then, to ensure consistency, the sum rule~\eqref{compress} is enforced when fitting $\mathrm{Im} g_{\rm w}(q,\omega)$. 

This procedure yielded $d = 1.3~\rm \AA$ with the Werder parameters and $d = 1.76~\rm \AA$ with the Aluru parameters. This last value agrees well with the position of the electronic density minimum at the water-graphene interface, as determined from DFT (main text Fig. 1b). This is consistent with the fact that the Aluru parameters are based on DFT calculations for water on a graphene surface~\cite{Wu2013}. Therefore, our sum-rule-based approach does place the origin of the $z$ axis where it would be expected from microscopic considerations.  

\begin{figure}
\centering
\vspace{-1cm}
\includegraphics{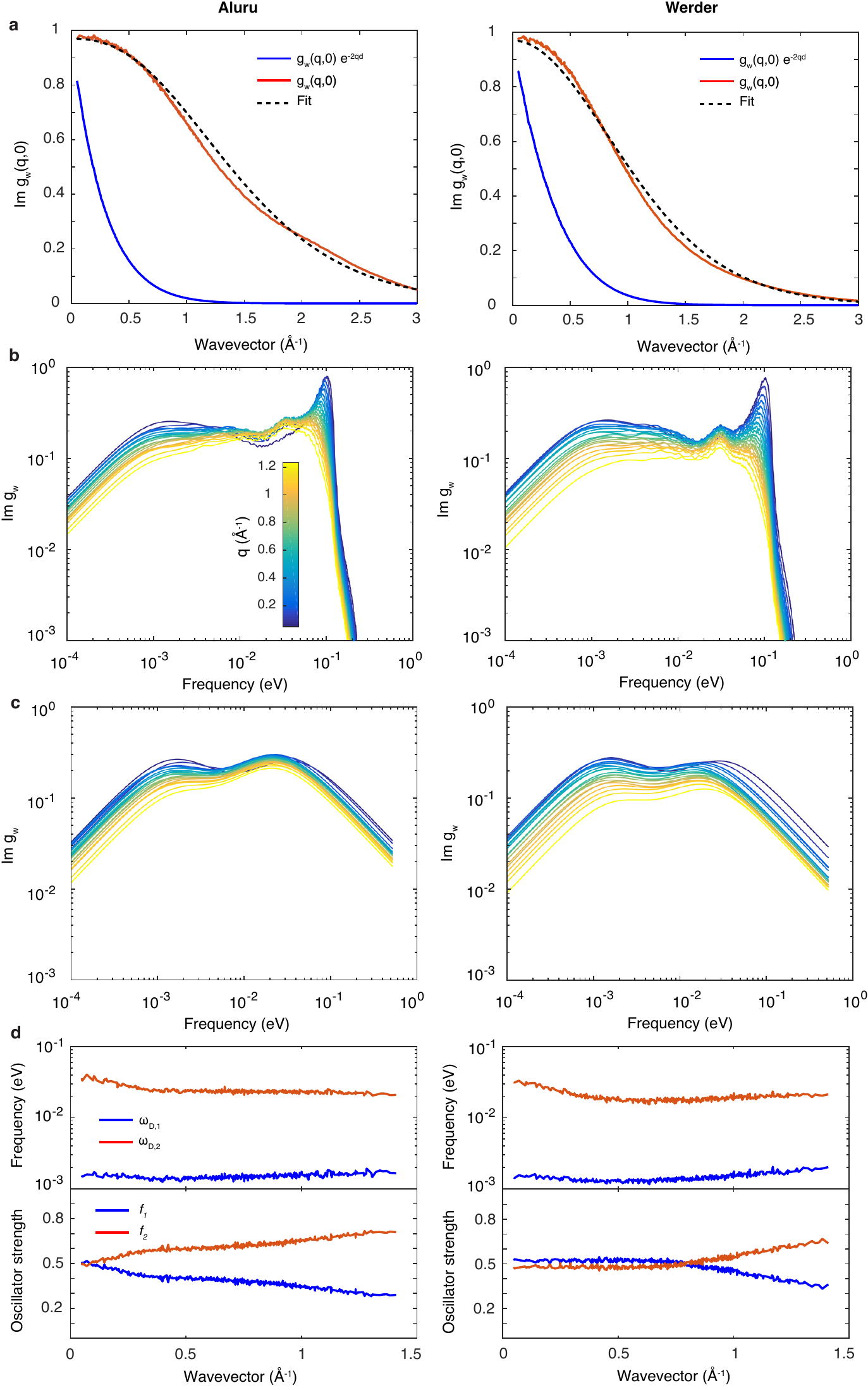}
\caption{\small{\textbf{Surface respose of water: simulation results}. The left (right) column corresponds to Aluru (Werder) parameters. \textbf{a}. Static surface response function, as obtained from the static structure factor, before and after cancellation of the gap $d$. The dashed line shows a fit to the data (eq.~\eqref{gqfit}). \textbf{b}. Surface response as a function of frequency, for different values of the wavevector $q$. \textbf{c}. Result of fitting the surface excitation spectra with two Debye peaks, for different wavevectors. \textbf{d}. Fitting parameter values. Frequencies of the Debye peaks (top), and relative oscillator strengths (bottom).}}
\end{figure}

\rev{\subsubsection{Slab simulations}
The slab simulations were analysed similarly to the interface simulations, with the gap $d$ set to the same value ($d = 1.3~\rm \AA$). The definition of the surface response function was generalised to a confined geometry: with the water slab lying between $z=0$ and $z=h$, we set 
\begin{equation}
g_{\rm w}(q,\omega) = - \frac{e^2}{4\pi \epsilon_0} \frac{2\pi}{q} \int_0^{h} \d z \d z' \, e^{-q(z+z')} \chi_{\rm w}(q,z,z',\omega). 
\end{equation}
The results are displayed in figure S2 for a range of wavevectors and slab thicknesses. Remarkably, the surface excitation spectra are found to depend very weakly on confinement: a difference with respect to the semi-infinite system is visible only for the smallest confinement $h = 1~\rm nm$. We attribute this to the relatively large values of the wavevectors $q$ under consideration: the effect of the confinement width $h$ is negligible if $e^{-2qh} \ll 1$. For wavevectors $q \gtrsim 0.2~\rm \AA^{-1}$ that are relevant for quantum friction, only the dynamics of the first water layer above the surface play a role. Therefore, our theory derived for two semi-infinite systems may be applied to planar nano-confined geometries. 

We note that in order to be applied to a nanotube geometry, our theory would in principle have to be extended to include the effect of surface curvature. Such an extension will be the subject of future work. However, curvature effects on the interfacial water structure are observed only in few-nanometre-radius nanotubes, and are negligible beyond 10~nm tube radius~\cite{Falk2010}. Therefore, we expect our semi-infinite theory to readily apply to the large radius ($15-50~\rm nm$) nanotubes such as those considered in the main text. }

\begin{figure}
\centering
\includegraphics{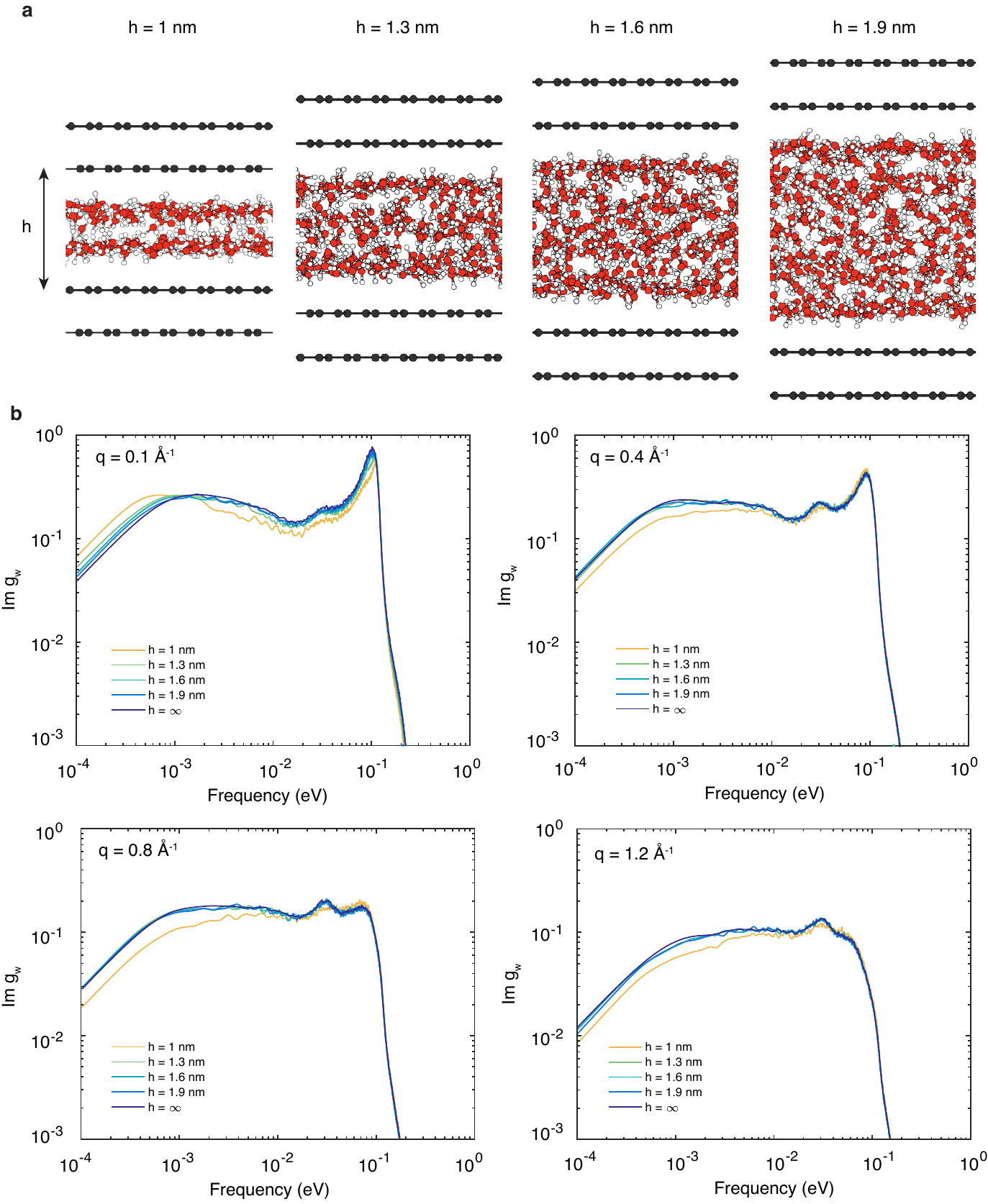}
\caption{\rev{\textbf{Surface respose of water: effect of confinement}. \textbf{a}. Simulation snapshots corresponding to the different confinement widths $h$. \textbf{b}. Surface excitation spectra $\mathrm{Im} \, g_{\rm w}(q,\omega)$ for a range of wavevectors $q$ and confinements $h$. }}
\end{figure}

\subsubsection*{Appendix}

The Kramers-Kr\"onig relation for the surface response function is 
\begin{equation}
\mathrm{Re} \, g_{\rm w}(q,\omega) =  \frac{1}{\pi} \mathcal{P} \int_{-\infty}^{+\infty} \d \omega' \frac{\mathrm{Im} g_{\rm w}(q,\omega')}{\omega'-\omega}. 
\end{equation}
Taking $\omega = 0$, this yields 
\begin{equation}
g_{\rm w}(q,0) = \frac{2}{\pi} \int_0^{+\infty}  \d \omega \, \frac{ \mathrm{Im} \, g_{\rm w}(q,\omega)}{\omega},
\end{equation}
since $ \mathrm{Im} \, g_{\rm w}(q,\omega)$ is an odd function of frequency. Using the fluctuation-dissipation theorem~\eqref{FDT}, we obtain 
\begin{equation}
\int_0^{+\infty}  \d \omega \, \frac{ \mathrm{Im} \, g_{\rm w}(q,\omega)}{\omega} = \frac{e^2}{8\epsilon_0 k_{\rm B} T} \int_{-\infty}^{+\infty} \d \omega \, \frac{S_{\rm s}(q,\omega)}{q}. 
\end{equation}
Then, using the definition~\eqref{structs} of the static structure factor, we recover the sum rule~\eqref{compress}. The surface structure factor is related to the full three-dimensional structure factor according to 
\begin{equation}
\bar S_{\rm s}(q) = \int_0^{+\infty} \d z \d z' e^{-q(z+z')} \bar S(q,z,z'),
\label{ssbulk}
\end{equation}
with $\bar S(q,z,z') = (1/\mathcal{A}) \langle n_{\rm w}(\q,z) n_{\rm w}(-\q,z') \rangle$. We may assume that the main contribution to the integral in eq.~\eqref{ssbulk} comes from the terms $z \approx z'$, due to the layering of the liquid near the surface~\cite{Barrat1999}. Then $S(q,z,z') \approx S(q|z) \delta (z-z')$, and we further assume that the structure factor is similar in all layers: $S(q|z) \approx S(q|0)$. Then, we have simply 
\begin{equation}
\bar S_{\rm s}(q) = \int_0^{+\infty} \d z e^{-2qz} \bar S(q|0) = \frac{S(q|0)}{2q},
\end{equation}
$S(q|0)$ is defined by the Fourier transform 
\begin{equation}
S(q|0) = \int \d \brho \, e^{-i \q \brho} S(\rho|0).
\label{SFourier}
\end{equation}
Since the structure factor is isotropic, the coefficients of odd powers of $q$ in a Taylor expansion around $q = 0$ vanish upon angular integration in eq.~\eqref{SFourier}. Hence the small $q$ expansion of $S_{\rm s}(q|0)$ contains only even powers of $q$, and the one of $S_{\rm s}(q)$ only odd powers, justifying the procedure described in section 4.2.2. 

\subsection{Fitting} 

The evaluation of quantum friction coefficients with eq.~\eqref{lambda} required to fit the MD simulation data, so as to avoid issues with numerical integration. The MD simulation results are reported in Fig. 2 of the main text, and figure S1. Figures S1 a and b show the static surface response function $g_{\rm w}(q,0)$, as obtained from the static structure factor (eq.~\eqref{compress}), before and after cancellation of the gap $d$, for both sets of LJ parameters. The simulation data was fitted with the following expression: 
\begin{equation}
g_{\rm w}(q,0) = \exp \left[ a + a'(1+(q/b)^\alpha)^{1/\alpha} \right], 
\label{gqfit}
\end{equation}
with the constraint $a + a' = \log(g_{\rm w}(0,0)) = \log(0.97)$, since the long wavelength limit of $g_{\rm w}(q,0)$ is imposed by the static permittivity $\epsilon(0) = 71$ of SPC/E water: $g_{\rm w}(0,0) = (\epsilon(0)-1)/(\epsilon(0)+1)$. The following results were obtained for the fit parameters:
\begin{center}
\begin{tabular}{|c|c|c|c|c|}
\hline
 & $a$ &$a'$ & $b~\rm (\AA^{-1})$ & $\alpha$  \\
\hline
 Werder & 5.16 & $-5.19$ & 1.95 & 2 \\
\hline
 Aluru & 3.38 & $-3.41$ & 1.79 & 2.4 \\
\hline 
\end{tabular}
\end{center}

The imaginary part of the surface response function was then fitted by a sum of two Debye peaks: 
\begin{equation}
\mathrm{Im} \, g_{\rm w}(q,\omega) = \mathrm{Im} \left[ \frac{f_1(q)}{1-i\omega/\omega_{D,1}(q)} +  \frac{f_2(q)}{1-i\omega/\omega_{D,2}(q)} \right], 
\end{equation}
with the constraint $f_1(q) + f_2(q) = g_{\rm w}(q,0)$ so as to satisfy the sum rule~\eqref{compress}. As can be seen in figure S1c, this fitting function reproduces quite well the general shape of the surface response function below $100~\rm meV$. The values we obtain for quantum friction coefficients are insensitive to the water response at higher frequencies, since these are cut off by the thermal factor in eq.~\eqref{lambda}. Furthermore, our classical simulations become inaccurate at frequencies above 100~meV, since at these frequencies the quantum nature of the dynamics can no longer be neglected (see, for instance, the comparison with an experimental result in the long wavelength limit, main text Fig. 2b). For both sets of LJ parameters, the frequencies of the two Debye peaks remain roughly constant (Fig. S1d), at $\omega_{D,1} = 1.5~\rm meV$ and $\omega_{D,2} = 20~\rm meV$. In the long wavelength limit, the two Debye peaks have the same oscillator strength, but the wavevector dependence is different for the two sets of parameters. The general trend is that the oscillator strength is transferred from the lower to the upper Debye peak as the wavevector increases. 

For evaluating quantum friction coefficients, we used the results obtained with Aluru parameters. Precisely, we represented the surface response function by the following analytical expression: 
\begin{equation}
g_{\rm w}(q,\omega) = \frac{g_{\rm w}(q,0)}{2} \left[ \frac{e^{-q/q_0}}{1-i\omega/\omega_{D,1}} +  \frac{2-e^{-q/q_0}}{1-i\omega/\omega_{D,2}}  \right],
\label{gwfinal}
\end{equation}
with $\omega_{D,1} = 1.5~\rm meV$, $\omega_{D,2} = 20~\rm meV$, $q_0 = 3.12~\rm \AA^{-1}$ and $g_{\rm w}(q,0)$ given by eq.~\eqref{gqfit}. The result in eq.~\eqref{gwfinal} is plotted in Fig. 2c of the main text as a function of frequency and wavevector. In Fig. 2b of the main text, the curve labeled "exp" is obtained from a fit to the experimentally determined dielectric function of bulk water reported in~\cite{Elton_thesis}.

\section{Jellium model}

\subsection{Surface response}
In order to qualitatively assess the role of electronic properties in the quantum friction of water, we consider a generic electronic system described within the infinite barrier jellium model, treated in the SR approximation (section 3.2). The non-interacting bulk density response function is computed according to eq.~\eqref{chi0bulk}. For the bulk jellium, there is a single band with energy $E(k) = \frac{\hbar^2 k^2}{2 m^*}$, with $m^*$ the effective mass, and the eigenstates are Bloch waves, so that all the matrix elements are equal to 1. Setting the temperature to 0, the Fermi-Dirac distribution becomes $f(E) = \theta(E_{\rm F} - E)$, with $E_{\rm F}$ the Fermi energy. The integral in eq.~\eqref{chi0bulk} may then be evaluated analytically, yielding the Lindhard function~\cite{Lindhard1954,bruus_ch13}: 
\begin{equation}
v(q,q_z) \mathrm{Re} [ \chi^0_{\rm B} (q,q_z,\omega)] = -\frac{\alpha r_{\rm s}}{4 x^2} \left( \frac{1}{2} + \frac{F(x,x_0) + F(x,-x_0)}{8x} \right),
\end{equation}
\begin{equation}
\begin{split}
v(q,q_z) \mathrm{Im} [ \chi^0_{\rm B} (q,q_z,\omega)] = -\frac{\alpha r_{\rm s}}{4 x^2} \left[ \frac{\pi(1-(x_0/x-x)^2)}{8x} \theta(x_0 - |x -x^2|) \theta(x + x^2-x_0) \dots \right. \\
+ \left.\frac{\pi x_0}{2 x} \theta(x_0) \theta(x-x^2-x_0) \right].
\end{split}
\end{equation}
We used the following notations: 
\begin{equation}
F(x,x_0) = \left( 1- \left( \frac{x_0}{x} -x \right)^2 \right) \, \log \left| \frac{x + x^2 -x_0}{x-x^2+x_0} \right|; 
\end{equation}
$x = \sqrt{q^2+q_z^2}/(2k_{\rm F})$, $x_0 = \omega/(4 \omega_{\rm F})$. $\omega_{\rm F} = E_{\rm F}/\hbar$ is the Fermi frequency, $k_{\rm F} = \sqrt{2 m^* E_{\rm F}/\hbar^2}$ is the Fermi wavevector. $v(q,q_z) = e^2/(\epsilon_0(q^2+q_z^2))$ is the 3D Coulomb potential. Once the frequencies and wavevectors have been normalised by the Fermi frequency and wavevector, the Lindhard function depends only on the electron density parameter $r_{\rm s}$, defined by 
\begin{equation}
r_{\rm s} = \left( \frac{9 \pi}{4} \right)^{1/3} \frac{(m^*/m_{\rm e})}{k_{\rm F} a_0},
\end{equation}
with $a_0 = 4\pi \epsilon_0 \hbar^2/(m_{\rm e} e^2)$ the Bohr radius and $m_{\rm e}$ the electron mass. The prefactor $\alpha$ is $\alpha = (4/\pi)(9\pi/4)^{-1/3} \approx 0.66$. Given the Lindhard function, the surface response function of the semi-infinite jellium can be evaluated by carrying out numerically the integration in eq.~\eqref{gspecular}. 

\subsection{Friction coefficient}

\begin{figure}
\centering
\includegraphics[scale=0.93]{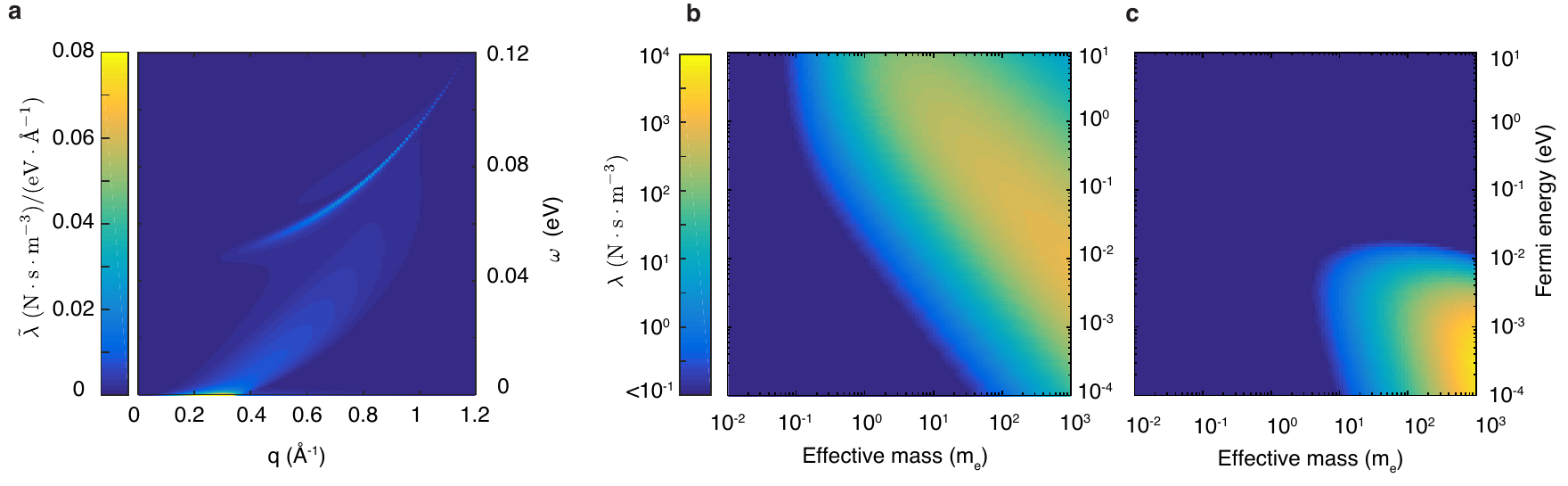}
\caption{\textbf{Quantum friction of water on a jellium}. \textbf{a}. The friction integrand $\tilde \lambda$ (defined in the text), for $E_{\rm F} = 2~\rm meV$ and $m^* = 50 m_{\rm e}$, as a function of $q$ and $\omega$. \textbf{b}. Low frequency particle-hole excitation contribution to the quantum friction coefficient. \textbf{c}. Surface plasmon contribution to the quantum friction coefficient.}
\end{figure}

In order to evaluate the quantum friction coefficient of water on a semi-infinite jellium according to eq. (6) of the main text, we fit the numerically determined jellium surface response function with analytical expressions. Such a procedure is necessary because the integral in eq. (7) has contributions at very low frequencies (typically, the water Debye mode frequency), and much higher frequencies corresponding to the electronic surface plasmon mode, which makes numerical sampling difficult. Figure S3a shows the integrand of eq. (7), that is 
\begin{equation}
\tilde \lambda(q,\omega) = \frac{\hbar^2}{8 \pi^2 k_B T}  \frac{q^3}{\mathrm{sinh}^{2}\left( \frac{\hbar \omega}{2k_B T} \right)} \frac{ \mathrm{Im}[g_{\rm e}(q,\omega)] \, \mathrm{Im}[g_{\rm w}(q,\omega)] }{|1-g_{\rm e}(q,\omega)\,g_{\rm w}(q,\omega)|^2},
\end{equation}
in $(q,\omega)$ space, with parameter values such that the surface plasmon frequency is low enough to make a non-negligible contribution ($E_{\rm F} = 2~\rm meV$ and $m^* = 50 \,m_{\rm e}$). We observe observe that the integrand $\tilde \lambda(q,\omega)$ has contributions from two disjoint regions of the $(q,\omega)$ space: at very low frequencies ($\omega \to 0$), and at frequencies around the surface plasmon frequency. Hence, we may fit the jellium surface response separately in these two regions. 

Our choice of fit functions is guided by~\cite{Penzar1984}. In the low frequency region, we use 
\begin{equation}
g_{\rm e}(q,\omega) = e^{- q/(k_{\rm F}C[r_{\rm s}])} + i A[r_{\rm s}] \frac{\omega}{E_{\rm F}} \frac{q}{k_{\rm F}} e^{-q/(k_{\rm F}B[r_{\rm s}])} \theta(q-2k_{\rm F}). 
\end{equation}
For the surface plasmon region, we fitted the imaginary part of the surface response function by a Lorentzian: 
\begin{equation}
\mathrm{Im}\, g_{\rm e}(q,\omega) = \frac{D[q/k_{\rm F},r_{\rm s}]}{2} \frac{1}{(\omega-\omega_P[q/k_{\rm F},r_{\rm s}])^2 + \gamma[q/k_{\rm F},r_{\rm s}]^2}. 
\end{equation}
The real part was then consistently determined through the Kramers-Kr\"onig relation: 
\begin{equation}
\mathrm{Re} \, g_{\rm e}(q,\omega) = \frac{2}{\pi} \mathcal{P} \int_0^{+\infty} \d \omega' \frac{\omega' \mathrm{Im}\, g_{\rm e}(q,\omega')}{\omega'^2-\omega^2} =  \frac{D[q/k_{\rm F},r_{\rm s}]}{2} \frac{\omega_P[q/k_{\rm F},r_{\rm s}](\omega_P[q/k_{\rm F},r_{\rm s}]-\omega)}{(\omega-\omega_P[q/k_{\rm F},r_{\rm s}])^2 + \gamma[q/k_{\rm F},r_{\rm s}]^2},
\label{KK}
\end{equation}
where $\mathcal{P}$ indicates that the Cauchy principal value of the integral is taken. We carry out the fit up to $q_{\rm max}/k_{\rm F} = -1+\sqrt{1+2 \omega_P^0[r_{\rm s}]/\omega_{\rm F}}$, where 
\begin{equation}
\omega_P^0[r_{\rm s}] = \sqrt{ \frac{8}{3\pi} \left(\frac{4}{9\pi}\right)^{1/3} } r_{\rm s}^{1/2} \omega_{\rm F} \approx 0.67 \times r_{\rm s}^{1/2} \omega_{\rm F}
\end{equation}
is the surface plasmon frequency at zero wavevector. For wavevectors beyond $q_{\rm max}$ the surface plasmon is Landau-damped, effectively disappearing from the spectrum. We found that the fit results were well-reproduced by the following analytical expressions: $A[r_{\rm s}] = 4.35 \times r_{\rm s}^{-0.9}$,
\begin{equation}
\omega_P[q,r_{\rm s}] = \omega_P^0[r_{\rm s}] \left(1 + \frac{q^2}{q_{\rm max}[r_{\rm s}]^2} \right) ~~~~ \mathrm{and} ~~~~ D[q,r_{\rm s}] = e^{-q/(0.45\times q_{\rm max}[r_{\rm s}])},
\end{equation}
and $\gamma = 0.01 \times \omega_P^0 [r_{\rm s}]$. The numerical values obtained for the parameters $B$ and $C$ are available on Zenodo~\cite{zenodo}. Figures S3 b and c show separately the low frequency and the plasmon contribution to the quantum friction coefficient as a function of the jellium parameters $m^*$ and $\omega_{\rm F}$. It clearly appears that the friction is dominated by the plasmon contribution in the region of low Fermi energy and high effective mass, where the plasmon is at low energy and is weakly damped up to high momenta. 

\subsection{Phonon contribution}

\begin{figure}
\centering
\includegraphics[scale=1]{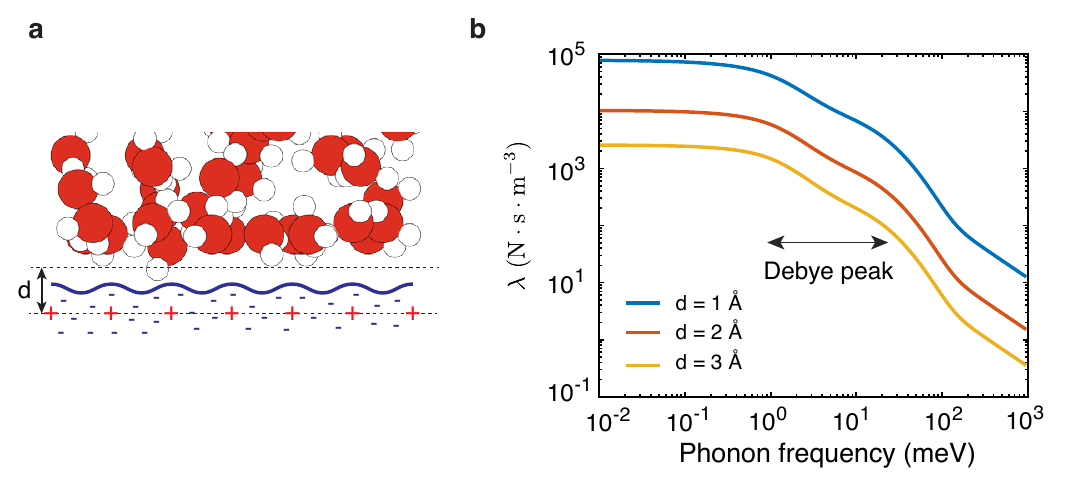}
\caption{\textbf{Phonon contribution to quantum friction}. \textbf{a}. Definition of the distance $d$ appearing in eq.~\eqref{lambdad}. \textbf{b}. Phonon contribution to the quantum friction coefficient as a function of phonon frequency $\omega_{\rm s}$, for different values of $d$. The phonon width is set at $\gamma = \omega_{\rm s}/20$.}
\end{figure}

In addition to electronic excitations, phonon modes can make a contribution to the surface response function of a solid, precisely in the low frequency and high momentum region relevant for water quantum friction. We expect that the phonon contribution will be most significant for polar materials such as hBN or $\rm SiO_2$, which have nearly dispersionless optical phonon modes, so that their dielectric response may be considered local up to momenta comparable with the Brillouin zone size. We consider for simplicity a material with a single optical phonon mode with frequency $\omega_{\rm ph}$ and width $\gamma$. Assuming $\epsilon_{\infty} \approx 1$, its dielectric function may be written as 
\begin{equation}
\epsilon(\omega) = \epsilon_{\infty} +(\epsilon_{\rm s} - 1) \frac{\omega_{\rm ph}^2}{\omega_{\rm ph}^2- \omega^2 - i \gamma \omega},
\end{equation}
where $\epsilon_{\infty}$ is the high frequency background dielectric constant and $\epsilon_{\rm s}$ is the static dielectric constant. The corresponding surface response function is
\begin{equation}
g(\omega) = \frac{\epsilon(\omega)-1}{\epsilon(\omega)+1} =  \frac{\epsilon_{\rm s}-1}{\epsilon_{\rm s}+1} \frac{\omega_{\rm s}^2}{\omega_{\rm s}^2 - \omega^2 - i \gamma \omega},
\label{gphonon}
\end{equation}
with $\omega_{\rm s} = \omega_{\rm ph} \sqrt{(1+\epsilon_{\rm s})/2}$ the surface phonon frequency. The ratio $(\epsilon_{\rm s} -1)/(\epsilon_{\rm s} + 1)$ defines a surface phonon "oscillator strength", which is close to 1 since $\epsilon_{\rm s} \sim 3 - 10$. 

Now, in the case of phonons, eq.~\eqref{gphonon} cannot be directly used to evaluate a quantum friction coefficient according to eq.~\eqref{lambda}. Indeed, one should take into account a microscopic distance $d$ between the interfacial water layer and the first atomic layer of the solid, which supports the phonon mode (see Fig. S4). Eq.~\eqref{lambda} is then modified according to 
\begin{equation}
\lambda^{\rm q} = \frac{\hbar^2}{8 \pi^2 k_B T} \int_0^{+\infty} \d q \, q^3 \color{blue} e^{-2 qd} \color{black} \int_{0}^{+\infty} \frac{\d \omega}{\mathrm{sinh}^{2}\left( \frac{\hbar \omega}{2k_B T} \right)} \frac{ \mathrm{Im}[g_{\rm e}^R(q,\omega)] \, \mathrm{Im}[g_{\rm w}^R(q,\omega)] }{|1-\color{blue} e^{-2qd} \color{black}g_{\rm e}^R(q,\omega)\,g_{\rm w}^R(q,\omega)|^2},
\label{lambdad}
\end{equation}
We note that in the case of electronic excitations, there is no need to add explicitly a distance $d$, as the water molecules are in direct contact with the solid's electronic density. The position of the image plane within the solid is then taken into account in the momentum dependence of the electronic surface response function. Figure S4 shows the friction coefficient computed according eq.~\eqref{lambdad} for different values of $d$ in the angstr\"om range as a function of phonon frequency (the "oscillator strength" is taken equal to 1). We assume a small phonon width $\gamma = \omega_{\rm s}/20$. We find that, for reasonable surface phonon frequencies ($\omega_{\rm s} \sim 100~\rm meV$), the phonon contribution to water quantum friction is rather small ($\lambda \sim 10^2~\rm N \cdot s \cdot m^{-3}$), even at $d = 1~\rm \AA$. 

In figure S4, we plotted the quantum friction coefficient for a wide range of phonon frequencies (beyond physically reasonable ones), in order to study in a simple case the dependence of the friction coefficient on mode frequencies. In eqs.~\eqref{lambda} and \eqref{lambdad}, the product of surface response functions suggests that there should be a resonance for the friction coefficient when a solid (phonon) mode has the same frequency as the water Debye mode. This is however not the case, and the friction coefficient is found to be a monotonically decreasing function of the phonon frequency. This is due to the thermal factor $\mathrm{sinh}(\hbar \omega/2k_{\rm B} T)$, which gives more weight to lower frequency modes, and to the very broad shape of the Debye peak. If the Debye peak was replaced by a weakly damped harmonic oscillator peak centred at $\omega_{\rm D}$, the friction coefficient would show a maximum at $\omega_{\rm s} = \omega_{\rm D}$, but would still converge to a non-zero value as $\omega_{\rm s} \to 0$. 

\section{Water-carbon interface}

\subsection{Graphene surface response function}

We compute the surface response function of monolayer graphene according to the definition in eq.~\eqref{gdef}. In order to explicit the density response function $\chi(q,z,z',\omega)$, we make use of the tight binding model of graphene~\cite{Blinowski1980,Shung1986}. In the tight binding description, the Hilbert space of the graphene $\pi$ electrons is restricted to linear combinations of localised states $|i\rangle$. We denote $c^{\dagger}_i, c_i$ the corresponding creation and annihilation operators and $\varphi(\rr-\rr_i)$ the corresponding wavefunctions. We use for $\varphi$ the generalised hydrogenic wavefunction representing the carbon $2p_z$ orbital~\cite{Shung1986}: 
\begin{equation}
\varphi (\rho,z) = A z e^{-Z \sqrt{\rho^2+z^2}/ 2a_0},
\label{2pz}
\end{equation}
with $A$ a normalisation factor, $a_0$ the Bohr radius and $Z = 3.18$. All overlaps between neighbouring orbitals are neglected, so that the electron density reads
\begin{equation}
n(\rr,t) = \sum_i |\varphi(\rr-\rr_i)|^2 c^{\dagger}_i(t) c_i (t) 
\label{grn}
\end{equation}
We introduce the Bloch state operators, defined separately for the sublattices $A$ and $B$: 
\begin{equation}
c^{\dagger}_A (\k) = \sum_{i \in A} e^{i \k \brho_i} c^{\dagger}_i ~~~~ \mathrm{and} ~~~~ c^{\dagger}_B(\k) = \sum_{i \in B} e^{i \k \brho_i} c^{\dagger}_i.  
\end{equation}
The inverse transformation is 
\begin{equation}
c^{\dagger}_{i \in A} = \int_{\rm BZ} \frac{\d \k}{\mathcal{A}_{\rm BZ}} e^{-i \k \brho_i} c^{\dagger}_{A} (\k) ~~~~ \mathrm{and} ~~~~ c^{\dagger}_{i \in B} = \int_{\rm BZ} \frac{\d \k}{\mathcal{A}_{\rm BZ}} e^{-i \k \brho_i} c^{\dagger}_{B} (\k),  
\label{bloch}
\end{equation}
with $\mathcal{A}_{\rm BZ}$ the area of the Brillouin zone. We further note the property
\begin{equation}
\sum_{i \in A,B} e^{i \k \brho_i} = \mathcal{A}_{\rm BZ} \sum_{\bf G} \delta (\k + \mathbf{G}),
\label{comb}
\end{equation}
where the $\bf G$ are vectors of the reciprocal lattice. Inserting \eqref{bloch} into \eqref{grn}, and using \eqref{comb}, we obtain 
\begin{equation}
n(\brho,z,t) = \int_{\rm BZ} \frac{\d \q}{(2 \pi)^2} \sum_{\bf G} \xi(\q + \GG,z) e^{i(\q + \GG) \brho}n_{\q}(t),
\end{equation}
with 
\begin{equation}
\xi(\q,z) = \int \d \brho |\varphi(\rho,z)|^2 e^{-i \q \brho}
\end{equation}
and 
\begin{equation}
n_{\q}(t) = \int_{\rm BZ} \frac{\d \k}{\mathcal{A}_{\rm BZ}} \sum_{a \in \{A,B\}} c^{\dagger}_a(\k+\q,t) c_a(\k,t). 
\label{nqgra}
\end{equation}
We now focus on the non-interacting density response function $\chi^0$. Inserting \eqref{nqgra} into the definition \eqref{chir}, we obtain 
\begin{equation}
\begin{split}
\chi^0(\rr,t,\rr',t') = \int_{\rm BZ} \frac{\d \q \d \q'}{(2\pi)^4} \sum_{\GG,\GG'} \xi(\q + \GG,z) \xi(\q' + \GG',z') e^{i(\q + \GG)\brho} e^{i (\q'+\GG')\brho'} \dots \\
\dots \left[ -\frac{i}{\hbar} \theta(t-t') \langle [n_{\q}(t),n_{\q'}(t')] \rangle_0\right]. 
\end{split} 
\label{chi01}
\end{equation}
Now, momentum conservation imposes 
\begin{equation}
\langle [n_{\q}(t),n_{\q'}(t')] \rangle_0 = \mathcal{A}_{\rm BZ} \delta( \q + \q') \langle [n_{\q}(t),n_{-\q}(t')] \rangle_0. 
\end{equation}
Hence, eq.~\eqref{chi01} becomes 
\begin{equation}
\begin{split}
\chi^0(\rr,t,\rr',t') = \int_{\rm BZ} \frac{\d \q }{(2\pi)^2} \sum_{\GG,\GG'} \xi(\q + \GG,z) \xi(\q + \GG',z')^* e^{i(\q + \GG)\brho} e^{-i (\q+\GG')\brho'} \dots \\
\dots \left[ -\frac{i}{\hbar} \theta(t-t') \frac{\mathcal{A}_{\rm BZ}}{(2\pi)^2}\langle [n_{\q}(t),n_{-\q}(t')] \rangle_0\right]. 
\end{split} 
\label{chi0GG}
\end{equation}
The quantity in brackets is the non-interacting density response function $\chi^0_{\rm 2D}$ of graphene that has been evaluated in the literature~\cite{Hwang2007,Wunsch2006}. It has the usual expression as a function of the graphene eigenenergies $E_{\nu} (\k)$ and eigenstates $|\k,\nu\rangle$, which is the 2D analogue of eq.~\eqref{chi0bulk}: 
\begin{equation}
\chi^0_{\rm 2D} (\q,\omega) =  \sum_{\nu,\nu' = \pm 1} \int_{\rm BZ} \frac{\mathrm{d}^2k}{2\pi^2} |\langle \mathbf{k} + \mathbf{q},\nu | e^{i \mathbf{q \cdot r}}| \mathbf{k} ,\nu'\rangle|^2\frac{n_{\rm F}[E_{\nu}(\mathbf{k+q})]-n_{\rm F}[E_{\nu'}(\mathbf{k})]}{E_{\nu}(\mathbf{k}+\mathbf{q})-E_{\nu'}(\mathbf{k})-\hbar (\omega +i\delta) }.
\label{chi0gr}
\end{equation}
For momenta $\q$ that are small compared to the intervalley distance ($1.7~\rm \AA^{-1}$), the integration over the Brillouin zone may be carried out separately for the two valleys, and the matrix elements have the expression 
\begin{equation}
 |\langle \mathbf{k} + \mathbf{q},\nu | e^{i \mathbf{q \cdot r}}| \mathbf{k} ,\nu'\rangle| = \frac{1}{2} \left( 1+ \nu \nu'\frac{k + q\cos \theta}{\lVert \k + \q \rVert} \right),
\end{equation}
where $\theta$ is the angle between $\k$ and $\q$. If one further assumes zero temperature, the integral in eq.~\eqref{chi0gr} can be carried out analytically, yielding the expression reported in~\cite{Wunsch2006}, which we reproduce here for completeness: 
\begin{equation}
\begin{split}
\chi^0_{\rm 2D} (q,\omega) =& -i\pi \frac{F(q,\omega)}{v_{\rm F}^2} - \frac{2 E_{\rm F}}{\pi v_{\rm F}^2} + \dots \\
&+ \frac{F(q,\omega)}{v_{\rm F}^2} \left[ G \left( \frac{\omega + 2 E_{\rm F}}{v_{\rm F} q} \right) - \theta \left( \frac{2 E_{\rm F} - \omega}{v_{\rm F} q}-1 \right) \left\{G \left( \frac{2 E_{\rm F} - \omega}{v_{\rm F} q} \right)- i \pi\right\} + \right. \dots \\
 &+\left. \theta\left(\frac{\omega- 2E_{\rm F}}{v_{\rm F} q}+1 \right) G \left( \frac{\omega-2E_{\rm F}}{v_{\rm F} q} \right) \right].
\end{split}
\end{equation}
Here, $E_{\rm F}$ is the graphene Fermi energy (doping level), $v_{\rm F} = 6.73~\rm eV\cdot \AA$ is the graphene Fermi velocity, and the functions $F$ and $G$ are given by 
\begin{equation}
F(q,\omega) = \frac{1}{4\pi} \frac{v_{\rm F}^2q^2}{\sqrt{\omega^2-v_{\rm F}^2q^2}},
\end{equation}
and
\begin{equation}
G(x) = x \sqrt{x^2-1} - \log \left( x + \sqrt{x^2-1} \right). 
\end{equation}

Having written out the non-interacting density response function for graphene, we may examine the corresponding surface response function. We found (eq.~\eqref{chi0GG}), that the non-interacting density response function $\chi^0$ has a Fourier expansion of the form
\begin{equation}
\chi^0(\rr,\rr',\omega) = \int_{\rm BZ} \frac{\d \q }{(2\pi)^2} \sum_{\GG,\GG'} \chi^0_{\GG \GG'} (\q,z,z',\omega). 
\end{equation}
Therefore, in principle, one should define a non-interacting surface response function for every $(\GG,\GG')$: 
\begin{equation}
g^0_{\GG \GG'} (\q,\omega) = -\frac{e^2}{2\epsilon_0 \lVert \q + \GG' \rVert} \int \d z \d z' e^{\lVert \q + \GG \rVert z} e^{\lVert \q + \GG' \rVert z'} \chi^0_{\GG \GG'} (\q,z,z',\omega). 
\label{g0gg}
\end{equation}
Physically, $g_{\GG \GG'}$ determines the induced potential at wavevector $\q + \GG'$ in response to an applied potential at wavevector $\q + \GG$. Now, since we are considering applied potentials at wavevectors much smaller than the reciprocal lattice spacing $G_1 = 2.9~\rm \AA^{-1}$, we will only be interested in $g^0_{00} (q,\omega) \equiv g^0(q,\omega)$. Expanding eq.~\eqref{g0gg}, 
\begin{equation}
g^0(q,\omega) =  -\frac{e^2}{2\epsilon_0 q} \int \d z \d z' e^{q z} e^{q z'}\xi(q,z) \xi(q,z')^* \chi^0_{\rm 2D}(q,\omega). 
\label{g00}
\end{equation}
In the water-graphene configuration, the $z$ integration should in principle run from $z = - \infty$ up to some fixed $z_0 >0$, which sets the limit between solid and liquid. However, the atomic orbitals $\varphi$ extend formally up to $z = +\infty$, and they should be somehow cutoff at $z = z_0$. We expect $z_0$ to be in the angstr\"om range, comparable to the typical extension of the $2p_z$ orbital. Hence, for wavevectors $q \lesssim 1/z_0$, the exact choice of $z_0$ plays no significant role. We may then set $e^{qz} \approx 1$ in eq.~\eqref{g00}, which reduces to 
\begin{equation}
g^0(q,\omega) = -\frac{e^2}{2\epsilon_0 q}|\xi(q)|^2 \chi^0_{\rm 2D}(q,\omega),
\end{equation}
with $\xi(q) \equiv \int \d z \xi(q,z)$. From the expression~\eqref{2pz} of the $2p_z$ orbital, one obtains~\cite{Shung1986}
\begin{equation}
|\xi(q)|^2 = \left(1+(qa_0/Z)^2\right)^{-6}. 
\end{equation}

In order to obtain the interacting surface response function, in general one has to solve the RPA Dyson equation~\eqref{dysonchi} for the density response function. However, for wavevectors smaller than the reciprocal lattice spacing, it reduces to a Dyson equation involving directly the surface response function: 
\begin{equation}
g(q,\omega) = g^0(q,\omega) - g^0(q,\omega) g(q,\omega). 
\label{dysong}
\end{equation}
Hence, we finally obtain the graphene surface response function as 
\begin{equation}
g(q,\omega) = \frac{-\dfrac{e^2}{2\epsilon_0 q}|\xi(q)|^2 \chi^0_{\rm 2D}(q,\omega)}{1-\dfrac{e^2}{2\epsilon_0 q}|\xi(q)|^2 \chi^0_{\rm 2D}(q,\omega)}. 
\end{equation}
This expression was used to obtain Fig. 4a and Fig. 4c of the main text. The charge carrier (electron or hole) density in graphene ($n$) is related to the Fermi level $E_{\rm F}$ according to $| E_{\rm F}| = v_{\rm F} \sqrt{\pi n}$.

\subsection{1D chain model}

In this section, we present a qualitative model for the surface response of graphite, which accounts for the presence of a dispersionless low energy mode in the surface excitation spectrum. It is based on the physical interpretation of this low energy mode as originating from interlayer excitations of electrons located mainly on the $B$ sublattices (main text Fig. 4b). The lack of dispersion indicates that, from the point of view of low energy excitations, the 1D chains formed by the B sublattice atoms behave as if they were independent, with no in-plane tunnelling between the chains. This behaviour can be tied, to some extent, to the flattening of the $\pi$ bands in graphite with respect to graphene, though the very flat dispersion observed in experiment does not seem to fully grasped by the band structure of graphite at the tight-binding level. Hence, as a phenomenological model for the experimentally observed graphite surface response, we consider an array of semi-infinite tight-binding chains, with coupling parameter $\gamma_2 = 10~\rm meV$~\cite{Partoens2006}. 

\subsubsection{Local Green's function}
As a first step, we evaluate the local non-interacting (retarded) Green's function at the topmost atom of a 1D chain, denoted $G_{11}(\omega)$. We introduce the creation and annihilation  operators $c^{\dagger}_i,c_i$ at the chain atoms, and the tight-binding Hamiltonian 
\begin{equation}
\hat H = \sum_{i = 1}^{\infty} \gamma_2 (c^{\dagger}_i c_{i+1} + c_i c^{\dagger}_{i+1}). 
\end{equation}
The Green's function $G_{11}$ is given by the first coefficient of the matrix $(\hbar \omega\hat I - \hat H)^{-1}$, with $\hat I$ the identity matrix: 
\begin{equation}
G_{11}(\omega) = (\hbar \omega\hat I - \hat H)^{-1}_{11}. 
\end{equation}
$G_{11}$ may thus be computed with the help of cofactor expansions. We denote $\hat H_N$ the Hamiltonian of a tight-binding chain with finite length $N$, and $\tilde H_N = \hbar \omega \hat I_N - \hat H_N$, so that the local Green's function at the finite chain's topmost atom is $\tilde G_{11}^N = (\tilde H_N^{-1})_{11}$. In matrix form, 
\begin{equation}
\tilde H_N = \left(
\begin{array}{ccccc}
\hbar \omega & -\gamma_2 & 0 & \ddots & \ddots \\
-\gamma_2 & \hbar \omega & - \gamma_2 &0 & \ddots \\
0  & -\gamma_2 & \hbar \omega & -\gamma_2 & \ddots \\
\ddots & 0 & -\gamma_2 & \hbar \omega & \ddots \\
\ddots & \ddots & \ddots & \ddots & \ddots \\
\end{array}
\right). 
\end{equation}
 Using the cofactor expansion formula for the matrix inverse, we obtain 
 \begin{equation}
 G^N_{11}(\omega) = \frac{ \det \tilde H_{N-1} }{\det \tilde H_N}. 
 \label{gn11}
 \end{equation}
 We then expand the determinant of $\tilde H_N$ along the first row: 
 \begin{equation}
 \det \tilde H_{N} = \hbar \omega \det \tilde H_{N-1} + \gamma_2 \det M_{N-1},
 \end{equation}
 with 
 \begin{equation}
 M_{N-1} = \left(
\begin{array}{ccccc}
-\gamma_2 & -\gamma_2 & 0 & \ddots & \ddots \\
0 & \hbar \omega & - \gamma_2 &0 & \ddots \\
0  & -\gamma_2 & \hbar \omega & -\gamma_2 & \ddots \\
\ddots & 0 & -\gamma_2 & \hbar \omega & \ddots \\
\ddots & \ddots & \ddots & \ddots & \ddots \\
\end{array}
\right). 
\end{equation}
Then, expanding the determinant of $ M_{N-1}$ along the first column, we find $\det  M_{N-1} = -\gamma_2 \det \tilde H_{N-2}$. Replacing into eq.~\eqref{gn11}, we obtain 
\begin{align}
 G^N_{11}(\omega) &= \frac{ \det \tilde H_{N-1} }{\hbar \omega \det \tilde H_{N-1} - \gamma_2^2\det \tilde H_{N-2}} \\
 & = \frac{1}{\hbar \omega - \gamma_2^2 G_{11}^{N-1}(\omega)}.
\end{align}
Taking the limit $N \to \infty$, this yields a self-consistent equation for $G_{11}$, which is solved by 
\begin{equation}
G_{11}(\omega) = \frac{1}{2 \gamma_2^2}\left( \omega - \sqrt{\omega^2-4 \gamma_2^2} \right). 
\end{equation}
Here we chose the $-$ sign in the solution of the second order equation so that $G_{11}$ has a negative imaginary part as required by causality. 

\subsubsection{Local density response}
We may now compute the non-interacting local density response function at the topmost atom, defined as
\begin{equation}
\chi_1 (\omega) = \int_{-\infty}^{+\infty} \d(t-t') e^{i\omega t} \left[ -\frac{i}{\hbar} \theta(t-t') \langle [n_1(t),n_1(t')] \rangle_0 \right],
\end{equation}
with $n_1 \equiv c^{\dagger}_1 c_1$. For that, we require the spectral function 
\begin{equation}
A_1(\omega) = -2 \,\mathrm{Im} \,G_{11}(\omega) = \frac{2}{\gamma_2} \sqrt{1-\left( \frac{\omega}{2 \gamma_2} \right)^2} \theta (2 \gamma_2 - |\omega|). 
\label{a1}
\end{equation}
Then, density response function is obtained as~\cite{rammer_ch6}
\begin{equation}
\chi_1(\omega) = \frac{1}{2 \pi^2} \int_{-\infty}^{+\infty} \d E \d E' \, \frac{n_{\rm F}(E')-n_{\rm F}(E)}{E'-E+\hbar \omega + i \delta} A_1(E) A_1(E'). 
\end{equation}
Using eq.~\eqref{a1},
\begin{equation}
\chi_1(\omega) = \frac{2}{\gamma \pi^2}  \int_{-1}^{+1} \d E \d E' \, \frac{n_{\rm F}(E')-n_{\rm F}(E)}{E'-E+\hbar \omega/(2\gamma_2) + i \delta} \sqrt{(1-E^2)(1-E'^2)} \equiv  \frac{2}{\gamma \pi^2}  \tilde \chi_1\left( \frac{\omega}{2\gamma_2} \right). 
\label{chitilde}
\end{equation}
The function $\tilde \chi_1(x)$ is evaluated numerically, assuming one electron per atom so that the Fermi level is at 0. In order to facilitate subsequent computations, we introduce an analytical representation for $\mathrm{Im}\, \tilde \chi_1(x)$ at $T = 300~\rm K$ and $x >0$: 
\begin{equation}
\mathrm{Im} \, \tilde \chi_1(x) = -0.38 \times x (2-x) \theta(2-x). 
\label{imchitilde}
\end{equation}
We then determine the corresponding real part through the Kramers-Kr\"onig relation (see eq.~\eqref{KK}):
\begin{equation}
\mathrm{Re} \, \tilde \chi_1(x) = 0.38 \times \frac{2}{\pi} \left(2 x \, \mathrm{Arctanh} \,\left[\tilde f_1(x/2)\right] -2 + \frac{1}{2} x^2 \log \left| \frac{4-x^2}{x^2} \right| \right),
\label{rechitilde}
\end{equation} 
with $\tilde f_1(x) = x \theta(1-x) + (1/x) \theta(x-1)$. The numerical result for $\chi_1(\omega)$ is plotted alongside the analytical representation in Fig. S5a. 

\subsubsection{Graphite surface response}

\begin{figure}
\centering
\includegraphics{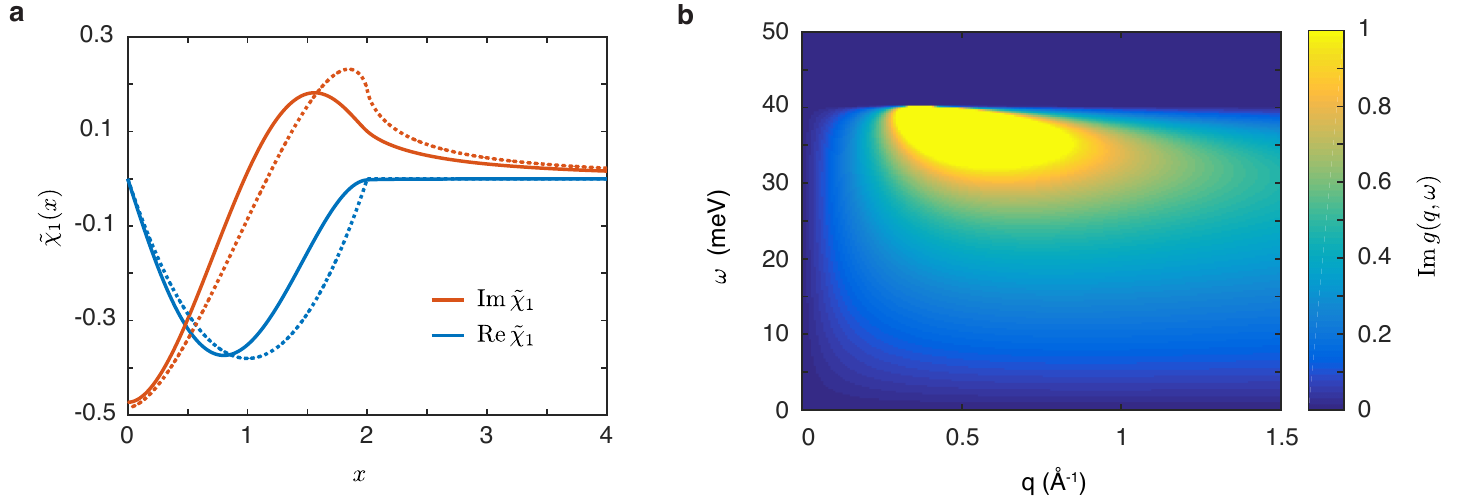}
\caption{\textbf{Independent chain model for the graphite surface response.} \textbf{a.} Normalised local density response function as defined in eq.~\eqref{chitilde}. Full lines are obtained by numerical integration, and the dashed lines correspond to the analytical representation in eqs.~\eqref{imchitilde} and \eqref{rechitilde}. \textbf{b.} Graphite surface response function as obtained in the independent chain model (eq.~\eqref{g1D}). We set $n_{\rm s} = 2.3\times 10^{12}~\rm cm^{-2}$ (free carrier density in graphite at 300 K). }
\end{figure}

We now use our results for a single atomic chain to obtain an expression for the surface response function of graphite, within our independent chain model. For not too small wavevectors ($q \gtrsim 1/(2c) = 0.14~\rm \AA^{-1}$), we may consider that the external field acts only on the first graphene layer. Then, the non-interacting density response function reads 
\begin{equation}
\chi^0(\rr,t,\rr',t') = \sum_{i,j \in B} | \varphi(\rr-\rr_i) |^2 |\varphi(\rr'- \rr_j)|^2 \left[ -\frac{i}{\hbar} \theta(t-t') \langle [n_i(t),n_j(t')] \rangle_0 \right]. 
\end{equation}
But since the 1D chains below atoms $i$ and $j$ are assumed to be decoupled (there is no electron tunnelling between them), $[n_i(t),n_j(t')] = \delta_{ij} [n_i(t),n_i(t')]$. Hence, 
\begin{equation}
\chi^0(\rr,t,\rr',t') = \sum_{i \in B} | \varphi(\rr-\rr_i) |^2 |\varphi(\rr'- \rr_i)|^2 \chi_1(\omega). 
\end{equation}
Carrying out Fourier transforms as in section 6.1, we obtain the Fourier components 
\begin{equation}
\chi^0_{\GG \GG'} (\q,z,z',\omega) = \frac{\mathcal{A}_{\rm BZ}}{(2\pi)^2} \chi_1(\omega) \xi(\q+\GG,z) \xi(\q+\GG',z')^*. 
\end{equation}
The factor $\mathcal{A}_{\rm BZ}/(2\pi)^2$ represents the electron density $n_{\rm s}$ in the first graphene layer, which is one electron per unit cell in our computation so far. However, we do not expect that all the $\pi$ electrons of the $B$ sublattice contribute to the low energy excitations we are describing, but rather only the free (electron and hole) charge carriers. Hence, from now on, we treat $n_{\rm s}$ as a parameter of our model, which represents the charge carrier density that contributes to the low energy interlayer excitations. 

Since we are making the approximation that the external field affects only atoms in the first graphene layer, we need only consider Coulomb interactions between atoms in that first layer. At wavevectors $q$ that are small enough to set $e^{qz} \approx 1$ for $z$ on the scale of a $p_z$ orbital, the interacting surface response function satisfies the same Dyson equation~\eqref{dysong} as for graphene. We then obtain 
\begin{equation}
g(q,\omega) = \frac{n_{\rm s} v_q |\xi(q)|^2 \chi_1(\omega)}{n_{\rm s} v_q |\xi(q)|^2 \chi_1(\omega)-1},
\label{g1D}
\end{equation}
with $v_q = e^2/(2\epsilon_0 q)$. This is equation (10) of the main text (where we have set $\chi_a(q,\omega) \equiv |\xi(q)|^2 \chi_1(\omega)$), which was used to obtain figures 4 d and e. We plot in figure S5b the imaginary of the surface response function in eq.~\eqref{g1D} as a function of $q$ and $\omega$, setting $n_{\rm s} = 2.3\times 10^{12}~\rm cm^{-2}$, which corresponds to the free carrier (electron and hole) density in graphite at 300 K\footnote{This value is inferred from the free carrier density in graphene at 0 K obtained from DFT calculations in \cite{Gruneis2008}, and from the values of plasma frequencies $\omega_{\rm P}$ at 0 K and 300 K: $n_{\rm s}[300~\mathrm{K}] = n_{\rm s}[0~\mathrm{K}]( \omega_{\rm P} [300~\mathrm{K}]^2/\omega_{\rm P} [0~\mathrm{K}]^2)$, with $n_{\rm s}[0~\mathrm{K}] = 10^{19}~\rm cm^{-2}$~\cite{Gruneis2008}, $\omega_{\rm P}(0~\mathrm{K}) = 19~\rm meV$ and $\omega_{\rm P} (300~\rm K) = 50~\rm meV$~\cite{Jensen1991}.} . We find that it accounts for a continuum of low energy excitations (below 40 meV), whose intensity decays slowly with increasing momentum, reproducing qualitatively the features observed in graphite electron energy loss spectroscopy~\cite{Laitenberger1996,Portail1999}. We note, however, that our model is bound to be incorrect at small values of $q$. Because of the 2D nature of our computation for the surface response function, its imaginary part decays to 0 at small $q$, while it would be expected to have a finite limit in a 3D setting. We therefore underestimate the friction coefficient by resorting to a 2D approximation, though we do not expect the underestimation to be significant, as the friction is most sensitive to the surface response function at large values of $q$.

\subsection{Multiwall nanotubes}

For the radius-dependence of the average interlayer spacing in multiwall carbon nanotubes, we rely on the transmission electron microscopy data of ref.~\cite{Endo1997}. For tube radii between 7.5 nm and 50 nm, we used a linear approximation to the data: 
\begin{equation}
d(R) = 3.35 + 0.002 \times (R\, (\mathrm{nm}) -50) ~~ (\rm \AA).
\end{equation}
In Fig. 4d of the main text, the curve labeled "Decoupling + gap" is obtained by considering a scaling $n_{\rm s}(R) = (1-p(R))n_{\rm s}^0 e^{-E_{\rm g}(R)/k_{\rm B}T}$.

\section{Density functional calculation}

We carried out an ab-initio molecular dynamics (AIMD) simulation of a graphene-water interface of area $12.83~\rm \AA \times 12.35~\rm \AA$ using the CP2K software. The simulation is identical to the one described in~\cite{Grosjean2019} with the difference that the hydroxide ion is replaced by a water molecule. A VASP calculation was performed on a single configuration extracted from the dynamics, again following~\cite{Grosjean2019}. We used the Perdew, Burke and Ernzerhof functional~\cite{Perdew1996} with the D3 dispersion correction scheme~\cite{Grimme2010,Grimme2011}, which has been shown to provide a good description of the water/graphene interface~\cite{Brandenburg2019}. Plane waves with kinetic energy larger than 600~eV were cut off and convergence was reached when the difference between total energy and eigenvalue energies was smaller than $10^{-6}~\rm eV$. The resulting electronic density, once averaged in the direction parallel to the interface, is shown in Fig. 1b of the main text.